\documentclass[jou,fignum]{apa} 

\usepackage{amssymb}
\usepackage{subfigure,graphicx,amstext}
\usepackage{amsthm,url}
\usepackage{color}

\title{Algorithmic complexity for psychology: A user-friendly implementation of the coding theorem method.}
\fourauthors{Nicolas Gauvrit}{Henrik Singmann}{Fernando Soler-Toscano}{Hector Zenil}
\fouraffiliations{CHArt (PARIS-reasoning), Ecole Pratique des Hautes Etudes, Paris, France}{Institut f\" ur Psychologie, Albert-Ludwigs-Universit\" at Freiburg, Freiburg, Germany.}{Grupo de L\'ogica, Lenguaje e Informaci\'on, Universidad de Sevilla, Spain.}{Unit of Computational Medicine, Center for Molecular Medicine, Karolinska Institute, Stockholm, Sweden}

\note{Corresponding author: Nicolas Gauvrit, ngauvrit@me.com}

\acknowledgements{Henrik Singmann is now at the Department of Psychology, University of Zurich (Switzerland).\\We would like to thank William Matthews for providing the data from his 2013 manuscript.}

\abstract{

Kolmogorov-Chaitin complexity has long been believed to be impossible to approximate when it comes to short sequences (e.g. of length 5-50). However, with the newly developed \emph{coding theorem method} the complexity of strings of length 2-11 can now be numerically estimated. We present the theoretical basis of algorithmic complexity for short strings (ACSS) and describe an R-package providing functions based on ACSS that will cover psychologists' needs and improve upon previous methods in three ways: (1) ACSS is now available not only for binary strings, but for strings based on up to 9 different symbols, (2) ACSS no longer requires time-consuming computing, and (3) a new approach based on ACSS gives access to an estimation of the complexity of strings of any length. Finally, three illustrative examples show how these tools can be applied to psychology. \\[.3cm]

}

\shorttitle{Complexity for psychology}

\begin{document}
\maketitle    
Randomness and complexity are two concepts which are intimately related and are both central to numerous recent developments in various fields, including finance \cite{taufemback2011,brandouy2012}, linguistics \cite{gruber2010,naranan2011}, neuropsychology \cite{machado2010,fernandez2011,fernandez2012}, psychiatry \cite{yang2012,takahashi2012}, genetics \cite{yagil2009,ryabko2013}, sociology \cite{elzinga2010} and the behavioral sciences \cite{watanabe2003,scafetta2009}. In psychology, randomness and complexity have recently attracted interest, following the realization that they could shed light on a diversity of previously undeciphered behaviors and mental processes. It has been found, for instance, that the subjective difficulty of a concept is directly related to its ``boolean complexity'', defined as the shortest logical description of a concept \cite{feldman2000,feldman2003,feldman2006}. In the same vein, visual detection of shapes has been shown to be related to contour complexity \cite{wilder2011}.

More generally, perceptual organization itself has been described as based on simplicity or, equivalently, likelihood \cite{chater1996,chater2003}, in a model reconciling the complexity approach (perception is organized to minimize complexity) and a probability approach (perception is organized to maximize likelihood), very much in line with our view in this paper. Even the perception of similarity may be viewed through the lens of (conditional) complexity \cite{hahn2003}.

Randomness and complexity also play an important role in modern approaches to selecting the ``best'' among a set of candidate models \cite<i.e., model selection; e.g.,>{myung_model_2006,kellen_recognition_2013}, as discussed in more detail below in the section called ``Relationship to complexity based model selection''.

Complexity can also shed light on short term memory storage and recall, more specifically, on the process underlying \emph{chunking}. It is well known that the short term memory span lies between 4 and 7 items/chunks  \cite{miller1956magical,cowan2001magical}. When instructed to memorize longer sequences of, for example, letters or numbers, individuals employ a strategy of subdividing the sequence into chunks \cite{baddeley1975word}. However, the way chunks are created remains largely unexplained. A plausible explanation might be that chunks are built via minimizing the complexity of each chunk. For instance, one could split the sequence ``AAABABABA'' into the two substrings ``AAA'' and ``BABABA''. Very much in line with this idea, Mathy and Feldman \citeyear{mathy2012} provided evidence for the hypothesis that chunks are units of ``maximally compressed code''. In the above situation, both ``AAA'' and ``BABABA'' are supposedly of low complexity, and the chunks tailored to minimize the length of the resulting compressed information.

Outside the psychology of short term memory, the complexity of pseudo-random human-generated sequences is related to the strength of executive functions, specifically to inhibition or sustained attention \cite{towse07}. In random generation tasks, participants are asked to generate random-looking sequences, usually involving numbers from 1 to 6 (dice task) or digits between 1 and 9. These tasks are easy to administer and very informative in terms of higher level cognitive abilities. They have been used in investigations in various areas of psychology, such as visual perception \cite{cardaci2009}, aesthetics \cite{boon2011}, development \cite{scibinetti2011,pureza2013}, sport \cite{audiffren2009}, creativity \cite{zabelina2012}, sleep \cite{heuer2005sleep,bianchi2013}, and obesity \cite{crova2013}, to mention a few. In neuropsychology, random number generation tasks and other measures of behavioral or brain activity complexity have been used to investigate different disorders, such as schizophrenia \cite{koike2011}, autism \cite{lai2010,maes2012,fournier2013}, depression \cite{fernandez2009}, PTSD \cite{pearson2011,curci2013}, ADHD \cite{sokunbi2013}, OCD \cite{bedard2009}, hemispheric neglect \cite{loetscher2009neglect}, aphasia \cite{Proios08}, and neurodegenerative syndromes such as Parkinson's and Alzheimer's disease \cite{brown1990,hahn2012}.

Perceived complexity and randomness are also of the utmost importance within the ``new paradigm psychology of reasoning'' \cite{over09}. As an example, let us consider the representativeness heuristic \cite{tversky1974judgment}. Participants usually believe that the sequence ``HTHTHTHT'' is less likely to occur than the sequence ``HTHHTHTT'' when a fair coin is tossed 8 times. This of course is mathematically wrong, since all sequences of 8 heads or tails, including these two, share the same probability of occurrence ($1/2^{8}$). The  ``old paradigm'' was concerned with finding such biases and attributing irrationality to individuals \cite{kahneman1982judgment}. In the ``new paradigm'' on the other hand, researchers try to discover the ways in which sound probabilistic or Bayesian reasoning can lead to the observed errors \cite{manktelow1993rationality,hahn_perceptions_2009}. 

We can find some kind of rationality behind the wrong answer, by assuming that individuals do not estimate the probability that a fair coin will produce a particular string $s$ but rather the ``inverse'' probability that the process underlying this string is mere chance. More formally, if we use $s$ to denote a given string, $R$ to denote the event where a string has been produced by a random process, and $D$ to denote  the complementary event where a string has been produced by a non-random (or deterministic) process, then individuals may assess $P(R|s)$ instead of $P(s|R)$. If they do so within the framework of formal probability theory (and the new paradigm postulates that individuals tend to do so), then their estimation of the probability should be such that Bayes' theorem holds:
\begin{equation} \label{eq:bayes}
P(R|s)=\frac{P(s|R)P(R)}{P(s|R)P(R)+P(s|D)P(D)}.
\end{equation}

Alternatively, we could assume that individuals do not estimate the complete inverse $P(R|s)$ but just the posterior odds of a given string being produced by a random rather than a deterministic process \cite{williams_why_2013}. Again, these odds are given by Bayes' theorem:
\begin{equation} \label{eq:post-odds}
\frac{P(R|s)}{P(D|s)}=\frac{P(s|R)}{P(s|D)}\times\frac{P(R)}{P(D)}.
\end{equation}
The important part of Equation~\ref{eq:post-odds} is the first term on the right-hand side, as it is a function of the observed string $s$ and independent of the prior odds $P(R)/P(D)$. This likelihood ratio, also known as the Bayes factor \cite{kass_bayes_1995}, quantifies the evidence brought by the string $s$ based on which the prior odds are changed. In other words, this part corresponds to the ``amount of evidence [$s$] provides in favor of a random generating process'' \cite{hsu2010subjective}. 

The numerator of the Bayes factor, $P(s|R)$, is easily computed, and amounts to $1/2^{8}$ in the example given above. However the other likelihood, the probability of $s$ given that it was produced by an (unknown) deterministic process, $P(s|D)$, is more problematic. Although this probability has been informally linked to complexity, to the best of our knowledge no formal account of that link has ever been provided in the psychological literature, although some authors have suggested such a link \cite<e.g.,>{chater1996}. As we will see, however, computer scientists have fruitfully addressed this question \cite{solomonoff1964a,solomonoff1964b,levin1974}. One can think of $P(s|D)$ as the probability that a randomly selected (deterministic) algorithm produces $s$. In this sense, $P(s|D)$ is none other than the so-called \emph{algorithmic probability of} $s$. This probability is formally linked to the algorithmic complexity $K(s)$ of the string $s$ by the following formula (see below for more details): 
\begin{equation} \label{eq:comp1}
K(s)\approx -\log_{2}(P(s|D)).
\end{equation}

A normative measure of complexity and a way to make sense of $P(s|D)$ are crucial to several areas of research in psychology. In our example concerning the representativeness heuristic, one could see some sort of rationality in the usually observed behavior if in fact the complexity of $s_1 = \mbox{``HTHTHTHT''}$ were lower than that of $s_2 = \mbox{``HTHHTHTT''}$ (which it is, as shown below). Then following Equation~\ref{eq:comp1}, $P(s_1|D) > P(s_2|D)$. Consequently, the Bayes factor for a string being produced by a random process would be larger for $s_2$ than for $s_1$. In other words, even when ignoring the question of the priors for a random versus deterministic process (which are inherently subjective and debatable) $s_2$ provides more evidence for a random process than $s_1$. 

Researchers have hitherto relied on an intuitive perception of complexity, or in the last decades developed and used several tailored measures of randomness or complexity  \cite{towse1998,barbasz2008estimate,schulter2010mittenecker,williams_why_2013,hahn_perceptions_2009} in the hope of approaching algorithmic complexity. Because all these measures rely upon choices that are partially subjective and each focuses on a single characteristic of chance, they have come under strong criticism \cite{gauvrit2014}. Among these measures, some have a sound mathematical basis, but focus on particular features of randomness. For that reason, contradictory results have been reported \cite{wagenaar1970,wiegersma1984}. The mathematical definition of complexity, known as Kolmogorov-Chaitin complexity theory \cite{kolmogorov1965,chaitin1966}, or simply algorithmic complexity, has been recognized as the best possible option by mathematicians \cite{li2008}  and psychologists \cite{griffiths2003,griffiths2004}. However, because algorithmic complexity was thought to be impossible to approximate for the short sequences we usually deal with in psychology (sequences of 5-50 symbols, for instance), it has seldom been used.

In this article, we will first briefly describe algorithmic complexity theory and its deep links with algorithmic probability (leading to a formal definition of the probability that an unknown deterministic process results in a particular observation $s$). We will then describe the practical limits of algorithmic complexity and present a means to overcome them, namely the coding theorem method, the root of \emph{algorithmic complexity for short strings} (ACSS). A new set of tools, bundled in a package for the statistical programming language \texttt{R} \cite{r2014} and based on ACSS, will then be described. Finally, three short applications will be presented for illustrative purposes.

\section{Algorithmic complexity for short strings}
\subsection{Algorithmic complexity}

As defined by Alan Turing, a universal Turing machine is an abstraction of a general-purpose computing device capable of running any computer program. Among universal Turing machines, some are \emph{prefix free}, meaning that they only accept programs finishing with an ``END'' symbol. 
The algorithmic complexity~\cite{kolmogorov1965,chaitin1966} -- also
called Kolmogorov or Kolmogorov-Chaitin complexity -- of a string $s$
is given by the length of the shortest computer program running on a
universal prefix-free Turing machine $U$ that produces $s$ and then
halts, or formally written, $K_U(s)=\min\{|p| : U(p)=s\}$.  From
the \emph{invariance theorem}~\cite{calude2002,li2008}, we know that
the impact of the choice of $U$ (that is, of a specific Turing
machine), is limited and independent of $s$. It means that for
  any other universal Turing machine $U'$, the absolute value of
  $K_U(s)-K_{U'}(s)$ is bounded by some constant $C_{U,U'}$ which
  depends on $U$ and $U'$ but not on the specific $s$. So $K(s)$ is
  usually written instead of $K_U(s)$.

More precisely, the invariance theorem states that $K(s)$ computed on
two different Turing machine will differ at most by an additive constant $c$, which is independent of $s$, but that can be arbitrary large. One consequence of this theorem is that there
actually are infinitely many different complexities, depending on the
Turing machine. Talking about ``the algorithmic complexity'' of a string is a
shortcut. This theorem also guarantees that asymptotically (or for
long strings), the choice of the Turing machine will have limited
impact. However, for short strings we are considering here, the impact
could be important, and different Turing machines can yield different values, or
even different orders. This limitation is not due to technical reasons,
but the fact that there is no objective default universal Turing machine one could pick to
compute ``the'' complexity of a string. As we will explain below, our
approach seeks to overcome this difficulty by defining what could be
thought of as a ``mean complexity'' as computed with random
Turing machines. Because we do not choose a particular Turing machine but sample the space of
all possible Turing machine (running on a blank tape), the result will be an objective estimation of algorithmic complexity, although it will of course differ from most specific $K_{U}$.

Algorithmic complexity gave birth to a definition of randomness. To put it in a nutshell, a string is random if it is complex (i.e. exhibit no structure). Among the most striking results of algorithmic complexity theory is the convergence in definitions of randomness. For example, using martingales, Schnorr \citeyear{schnorr1973} proved that Kolmogorov random (complex) sequences are effectively unpredictable and vice versa; Chaitin~\citeyear{chaitin2004}  proved that Kolmogorov random sequences pass all effective statistical tests for randomness and vice versa, and are therefore equivalent to Martin-L\"{o}f randomness~\cite{martin1966}, hence the general acceptance of this measure. $K$ has become the accepted ultimate universal definition of complexity and randomness in mathematics and computer science~\cite{downey2008, nies2009, zenil2011book}.

One generally offered caveat regarding $K$ is that it is \emph{uncomputable}, meaning there is no Turing machine or algorithm that, given a string $s$, can output $K(s)$, the length of the shortest computer program $p$ that produces $s$. In fact, the theory shows that no computable measure can be a universal complexity measure. However, it is often overlooked that $K$ is \emph{upper semi-computable}. This means that it can be effectively approximated from above. That is, that there are effective algorithms, such as lossless compression algorithms, that can find programs (the decompressor $+$ the data to reproduce the string $s$ in full) giving upper bounds of Kolmogorov complexity. However, these methods are inapplicable to short strings (of length below 100), which is why they are seldom used in psychology. One reason why compression algorithms are unadapted to short strings
is that compressed files include not only the instructions to
  decompress the string, but also file headers and other data
  structures. It makes that for a short string $s$, the size of a
  compressed text file with just $s$ is longer than $|s|$. Another
reason is that, as it is the case with the complexity $K_U$
  associated with a particular Turing machine $U$, the choice of the
compression algorithm is crucial for short strings, because of
  the invariance theorem.

\subsection{Algorithmic probability and its relationship to algorithmic complexity}

A universal prefix-free Turing machine $U$, can also be used to define a probability measure $m$ on the set of all possible strings by setting $m(s)=\sum_{p:U(p)=s} 1/2^{|p|}$, where $p$ is a program of length $|p|$, and $U(p)$ the string produced by the Turing machine $U$ fed with program $p$. The Kraft inequality~\cite{calude2002} guarantees that $0\leq \sum_{s} m(s) < 1$. The number $m(s)$ is the probability that a randomly selected deterministic program will produce $s$ and then halt, or simply the \emph{algorithmic probability of $s$,} and provides a formal definition of $P(s|D)$, where $D$ stands for a generic deterministic algorithm \cite<for a more detailed description see>{gauvrit2014}. Numerical approximations to
$m(s)$ using standard Turing machines have shed light on the stability
and robustness of $m(s)$ in the face of changes in $U$, providing
examples of applications to various areas leading to semantic
measures~\cite{cilibrasi2005,cilibrasi2007}, which today are accepted
as regular methods in areas of computer science and linguistics, to
mention but two disciplines.

Recent work~\cite{delahaye2012,zenil2011thesis} has suggested that  approximating $m(s)$ could in practice be used to approximate $K$ for short strings. Indeed, the \emph{algorithmic coding theorem}~\cite{levin1974} establishes the connection as $K(s) =-\log_2 m(s)+O(1)$, where $O(1)$ is bounded independently of $s$. This relationship shows that strings with low $K(s)$ have the highest probability $m(s)$, while strings with large $K(s)$ have a correspondingly low probability $m(s)$ of being generated by a randomly selected deterministic program. This approach, based upon and motivated by algorithmic probability, allows us to approximate $K$ by means other than lossless compression, and has been recently applied  to financial time series ~\cite{brandouy2012, zenil2011} and in psychology~\cite<e.g.>{gauvrit2014vision}. The approach is equivalent to finding the best possible
  compression algorithm with a particular computer program
  enumeration.

Here, we extend a previous method addressing the question of binary strings' complexity~\cite{gauvrit2014} in several ways. First, we provide a method to estimate the complexity of strings based on any number of different symbols, up to 9 symbols. Second, we provide a fast and user-friendly algorithm to compute this estimation. Third, we also provide a method to approximate the complexity (or the local complexity) of strings of medium length (see below).

\subsection{The invariance theorem and the problem with short 
    strings}
\label{sec:ferthe-invar-theor}

The invariance theorem does not provide a reason to expect $-\log_2(m(s))$ and $K(s)$ to induce the same ordering over short strings. Here, we have chosen a simple and standard Turing machine model \cite<the Busy Beaver model;>{rado1962}\footnote{A demonstration is available online at \texttt{http://demonstrations.wolfram.com/BusyBeaver/}} in order to build an output distribution based on the seminal concept of algorithmic probability. This output distribution then serves as an objective complexity measure producing results in agreement both with intuition and with $K(s)$ to which it will converge for long strings guaranteed by the invariance theorem.
Furthermore, we have found that estimates of $K(s)$ are strongly correlated to those produced by lossless compression algorithms as they have traditionally been used as estimators of $K(s)$ (compressed data is a sufficient test of non-randomness, hence of low $K(s)$), when both techniques (the coding theorem method and lossless compression) overlap in their range of application for medium size strings in the order of hundreds to 1K bits.

The lack of guarantee in obtaining $K(s)$ from $-\log_2(m(s))$ is
  a problem also found in the most traditional method to estimate
  $K(s)$. Indeed, there is no guarantee that some lossless compressor
  algorithm will be able to compress a string that is compressible
  by some (or many) other(s). We do have statistical evidence that, at
  least with the Busy Beaver model, extending or reducing the
  Turing machine sample space does not impact the order 
  \cite{Kolmo2D,soler2012,soler2013}. We also found a correlation in
  output distribution using very different computational formalisms,
  e.g. cellular automata and Post tag systems \cite{AlgNatZenil}. We have also shown \cite{Kolmo2D} that $-\log_2(m(s))$
produces results compatible with compression methods 
 $K(s)$, and strongly correlates to direct $K(s)$
calculation \cite<length of first shortest Turing machine found producing
$s$;>{soler2013}.

\subsection{The coding theorem method in practice}

The basic idea at the root of the coding theorem method is to compute an approximation of $m(s)$. Instead of choosing a particular Turing machine, we ran a huge sample of Turing machines, and saved the resulting strings. The distribution of resulting strings gives the probability that a randomly selected Turing machine (equivalent to a universal Turing machine with a randomly selected program) will result in a given string. It therefore approximates $m(s)$. From this, we estimate an approximation $K$ of the algorithmic complexity of any string $s$ using the equation $K(s)=-\log_{2}(m(s))$. To put it in a nutshell again: A string is complex and henceforth random if the likelihood of it being produced by a randomly selected algorithm is low, which we estimated as described in the following.

To actually build the frequency distributions of strings with different numbers of symbols, we used a Turing machine simulator, written in \texttt{C++}, running on a supercomputer of middle-size at CICA (Centro Inform\'atico Cient\'ifico de Andaluc\'ia). The simulator runs Turing machines in $(n,m)$ ($n$ is the number of states of the Turing machine, and $m$ the number of symbols it uses) over a blank tape and stores the output of halting computations. For the generation of random machines, we used the implementation of the Mersenne Twister in the Boost \texttt{C++} library.

\begin{table}[t!]
\centering
\begin{tabular}{cccc}
  \hline
$(n,m)$ & Steps & Machines & Time \\
  \hline
(5,2) & 500 &  $9\,658\,153\,742\,336$  &
      450 days \\
      (4,4) & 2000 & $3.34 \times 10^{11}$ & 62 days \\
      (4,5) & 2000 & $2.14\times 10^{11}$  & 44 days \\
      (4,6) & 2000 & $1.8\times 10^{11}$  & 41 days\\
      (4,9) & 4000 & $2\times 10^{11}$ & 75 days  \\
   \hline
\end{tabular}
\caption{Data of the computations to build the frequency distributions}
\label{tab:samplesData}
\end{table}
  
\begin{figure}[b]
  \centering
    \includegraphics[width=.45\textwidth]{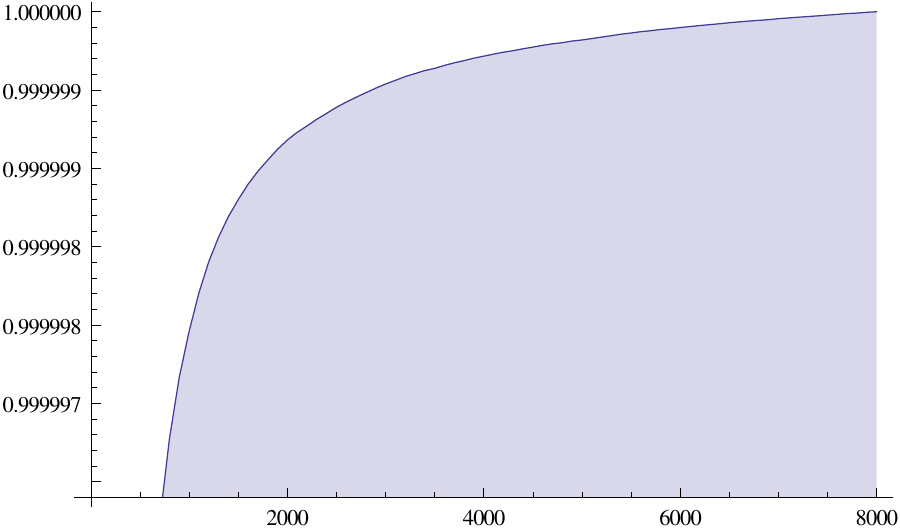}
  \caption{Runtime distribution in $(4,4)$. More than 99.9999\% of the Turing machines that stopped in 8000 steps or less actually halted before 2000 steps.}
  \label{fig:runtime44}
\end{figure}

Table~\ref{tab:samplesData} summarizes the size of the computations to build the distributions. The data corresponding to $(5,2)$ comes from a full
exploration of the space of Turing machines with 5 states and 2
symbols, as explained in Soler-Toscano, Zenil, Delahaye and Gauvrit~\citeyear{soler2012,soler2013}. All other
data, previously unpublished, correspond to samples of machines. The second column is the runtime cut. As the detection of non-halting machines is an undecidable problem, we stopped the computations exceeding that runtime. To determine the runtime bound, we first picked a sample of machines with an apt runtime $T$. For example, in the case of $(4,4)$, we ran $1.68 \times 10^{10}$ machines with a runtime cut of 8000 steps. For halting machines in that sample, we built the runtime distribution (Fig.~\ref{fig:runtime44}). Then we chose a runtime lower than $T$ with an accumulated halting probability very close to 1. That way we chose 2000 steps for
$(4,4)$. In Soler-Toscano et al.~\citeyear{soler2012} we argued that, following this methodology, we were able to cover the vast majority of halting machines.

The third column in Table~\ref{tab:samplesData} is the size of the
sample, that is, the machines that were actually run at the \texttt{C++}
simulator. After these computations, several symmetric completions were
applied to the data, so in fact the number of machines represented in
the samples is greater. For example, we only considered machines
moving to the right at the initial transition. So we complemented the set
of output strings with their reversals. More details about the
shortcuts to reduce the computations and the completions can be found
elsewhere~\cite{soler2013}. The last column in Table~\ref{tab:samplesData} is an
estimate of the time the computation would have taken on a
single processor at the CICA supercomputer. As we used between 10 and 70 processors, the actual times the computations took were shorter. In the following paragraphs we offer some details about the datasets obtained for each set of symbols. 

\subsubsection{(5,2)}
\label{sec:5-2}

This distribution is the only one previously
published~\cite{soler2012,gauvrit2014}. It consists of 99\,608 different binary
strings. All strings up to length 11 are included, and only 2 strings
of length 12 are missing.

\subsubsection{(4,4)}
\label{sec:4-4}

After applying the symmetric completions to the $3.34\times 10^{11}$ machines in the sample, we obtained a dataset representing the
output of $325\,433\,427\,739$ halting machines producing
$17\,768\,208$ different string patterns\footnote{Two strings correspond to the same pattern if one is obtained from the other by changing the symbols, such as ``11233'' and ``22311''. The pattern is the structure of the string, in this example described as ``one symbol repeated once, a second one, and a third one repeated once''. Given the definition of Turing Machines, strings with the same patterns always share the same complexity.}. To
reduce the file size and make it usable in practice, we selected only those patterns with 5 or more occurrences, resulting in a total of $1\,759\,364$ string patterns. In the final dataset, all strings comprising 4 symbols up to length 11 are represented by these patterns.

\subsubsection{(4,5)}
\label{sec:4-5}

After applying the symmetric completions, we obtained a dataset
corresponding to the output of $220\,037\,859\,595$ halting
machines producing $39\,057\,551$
different string patterns. Again, we selected only those patterns with 5 or more
occurrences,  resulting in a total of $3\,234\,430$ string patterns. In the final dataset, all strings comprising 5 symbols up to length 10 are represented by these patterns. 

\subsubsection{(4,6)}
\label{sec:4-6}

After applying the symmetric completions,we obtained a dataset
corresponding to the output of $192\,776\,974\,234$ halting machines
producing $66\,421\,783$ different string patterns. Here we selected only
those patterns with 10 or more occurrences, resulting in a total of $2\,638\,374$ string
patterns. In the final dataset, all strings with 6 symbols up to length 10 are represented
by these patterns.

\subsubsection{(4,9)}
\label{sec:4-9}

After applying the symmetric completions, we obtained
$231\,250\,483\,485$ halting Turing machines  
producing $165\,305\,964$
different string patterns. We selected only those with 10 or more occurrences, resulting in a total of $5\,127\,061$ string patterns. In the final dataset, all strings comprising 9 symbols up to length 10 are represented by these
patterns.

\subsubsection{Summary}
We approximated the algorithmic complexity of short strings (ACSS) using the coding theorem method by running huge numbers of randomly selected Turing machines (or all possible Turing machines for strings of only 2 symbols) and recorded their halting state. The resulting distribution of strings obtained approximates the complexity of each string; a string that was produced by more Turing machines is less complex (or random) than one produced by fewer Turing machines. The results of these computations, one dataset per number of symbols (2, 4, 5, 6, and 9), were bundled and made freely available. Hence, although the initial computation took weeks, the distributions are now readily available for all researchers interested in a formal and mathematically sound measure of the  complexity of short strings.

\section{The \texttt{acss} packages}

To make ACSS available, we have released two packages for the statistical programming language \texttt{R} \cite{r2014} under the GPL license \cite{gplv3} which are available at the Central R Archive Network (CRAN; \url{http://cran.r-project.org/}). An introduction to \texttt{R} is, however, beyond the scope of this manuscript. We recommend the use of \texttt{RStudio} (\url{http://www.rstudio.com/}) and refer the interested reader to more comprehensive literature \cite<e.g.,>{jones_introduction_2009,maindonald_data_2010,matloff_art_2011}.

The first package, \texttt{acss.data}, contains only the calculated datasets  described in the previous section in compressed form (the total size is 13.9 MB) and should not be used directly. The second package, \texttt{acss}, contains (a) functions to access the datasets and obtain ACSS and (b) functions to calculate other measures of complexity, and is intended to be used by researchers in psychology who wish to analyze short or medium-length (pseudo-)random strings. When installing or loading \texttt{acss} the data-only package is automatically installed or loaded. To install both packages, simply run \texttt{install.packages("acss")} at the \texttt{R} prompt. After installation, the packages can be loaded with \texttt{library("acss")}.\footnote{More information is  available at \url{http://cran.r-project.org/package=acss} and the documentation for all functions is available at \url{http://cran.r-project.org/web/packages/acss/acss.pdf}.} The next section describes the most important functions currently included in the package.

\subsection{Main functions}
All functions within \texttt{acss} have some common features. Most importantly, the first argument to all functions is \texttt{string}, corresponding to the string or strings for which one wishes to obtain the complexity measure. This argument necessarily needs to be a character vector (to avoid issues stemming from automatic coercion). In accordance with \texttt{R}'s general design, all functions are fully vectorized, hence \texttt{string} can be of length $> 1$ and will return an object of corresponding size. In addition, all functions return a named object, the names of the returned objects corresponding to the initial strings.

\subsubsection{Algorithmic complexity and probability}
The main objective of the \texttt{acss} package is to implement a convenient version of algorithmic complexity and algorithmic probability for short strings. The function \texttt{acss()} returns the ACSS approximation of the complexity $K(s)$ of a string $s$ of length between 2 and 12 characters, based on alphabets with either 2, 4, 5, 6, or 9 symbols, which we shall hereafter call $K_{2},K_{4},K_{5},K_{6}$ and $K_{9}$. The result is thus an approximation of the length of the shortest program running on a Turing machine that would produce the string and then halt. 

The function \texttt{acss()} also returns the observed probability $D(s)$ that a string $s$ of length up to 12 was produced by a randomly selected deterministic Turing machine. Just like $K$, it may be based on alphabets of 2, 4, 5, 6 or 9 symbols, hereafter called $D_{2}, D_{4},  ..., D_{9}$. As a consequence of returning both $K$ and $D$, \texttt{acss()} per default returns a matrix \footnote{Remember (e.g., Equation~\ref{eq:comp1}) that the measures $D$ and $K$ are linked by simple relations derived from the coding theorem method : $$K(s)=-\text{log}_{2}(D(s))\qquad D(s)=2^{-K(s)}.$$}. Note that the first time a function accessing \texttt{acss.data} is called within a \texttt{R} session, such as \texttt{acss()}, the complete data of all strings is loaded into the RAM which takes some time (even on modern computers this can take more than 10 seconds).
\begin{verbatim}
> acss(c("aba", "aaa"))
         K.9          D.9
aba 11.90539 0.0002606874
aaa 11.66997 0.0003068947
\end{verbatim}

Per default, \texttt{acss()} returns $K_9$ and $D_9$, which can be used for strings up to an alphabet of 9 symbols. To obtain values from a different alphabet one can use the \texttt{alphabet} argument (which is the second argument to all \texttt{acss} functions), which for \texttt{acss()} also accepts a vector of length $> 1$. If a string has more symbols than are available in the alphabet, these strings will be ignored:
\small
\begin{verbatim}
> acss(c("01011100", "00030101"), alphabet = c(2, 4))
              K.2      K.4          D.2          D.4
01011100 22.00301 24.75269 2.379222e-07 3.537500e-08
00030101       NA 24.92399           NA 3.141466e-08
\end{verbatim}
\normalsize

\subsubsection{Local complexity}

When asked to judge the randomness of sequences of medium length (say 10-100) or asked to produce pseudo-random sequences of this length, the limit of human working memory becomes problematic, and individuals likely try to overcome it by subdividing and performing the task on subsequences, for example, by maximizing the local complexity of the string or averaging across local complexities \cite{hahn_experiential_2014,hahn_perceptions_2009}. This feature can be assessed via the \texttt{local\_complexity()} function, which returns the complexity of substrings of a string as computed with a sliding window of substrings of length \texttt{span}, which may range from 2 to 12. The result of the function as applied to a string $s=s_{1}s_{2}\ldots s_{l}$ of length $l$ with a span $k$ is a vector of $l-k+1$ values. The $i$-th value is the complexity (ACSS) of the substring $s_{[i]}=s_{i}s_{i+1}\ldots s_{i+k-1}$. 

As an illustration, let's consider the 8-character long string based on a 4-symbol alphabet, ``aaabcbad''. The local complexity of this string with \texttt{span} $= 6$ will return $K_{4}(\text{aaabcb})$, $K_{4}(\text{aabcba})$, and $K_{4}(\text{abcbad})$, which equals  $(18.6, 19.4, 19.7)$:
\small 
\begin{verbatim}
> local_complexity("aaabcbad", alphabet=4, span=6)
$aaabcbad
  aaabcb   aabcba   abcbad 
18.60230 19.41826 19.71587 
\end{verbatim}
\normalsize

\subsubsection{Bayesian approach}

As discussed in the introduction, complexity may play a crucial role when combined with Bayes' theorem \cite<see also>{williams_why_2013,hsu2010subjective}. Instead of the observed probability $D$ we actually may be interested in the likelihood of a string $s$ of length $l$ given a deterministic process $P(s|D)$. As discussed before, this likelihood is trivial for a random process and amounts to $P(s|R) = 1/m^l$, where $m$, as above, is the size of the alphabet. To facilitate the corresponding usage of ACSS, \texttt{likelihood\_d()} returns the likelihood $P(s|D)$, given the actual length of $s$. This is done by taking $D(s)$ and normalizing it with the sum of all $D(s_i)$ for all $s_i$ with the same length $l$ as $s$ (note that this also entails performing the symmetric completions). As expected in the beginning, the likelihood of ``HTHTHTHT'' is larger than that of  ``HTHHTHTT'' under a deterministic process:

\footnotesize
\begin{verbatim}
> likelihood_d(c("HTHTHTHT","HTHHTHTT"), alphabet=2)
   HTHTHTHT    HTHHTHTT 
0.010366951 0.003102718 
\end{verbatim}
\normalsize

With the likelihood at hand, we can make full use of Bayes' theorem, and \texttt{acss} contains the corresponding functions. One can either obtain the likelihood ratio (Bayes factor) for a random rather than deterministic process via function \texttt{likelihood\_ratio()}. Or, if one is willing to make assumptions on the prior probability with which a random rather than deterministic process is responsible (i.e., $P(R) = 1-P(D)$) one can obtain the posterior probability for a random process given $s$, $P(R|s)$, using \texttt{prob\_random()}. The default for the prior is $P(R) = 0.5$. 

\footnotesize
\begin{verbatim}
> likelihood_ratio(c("HTHTHTHT", "HTHHTHTT"),
	alphabet = 2)
 HTHTHTHT  HTHHTHTT 
0.3767983 1.2589769 
> prob_random(c("HTHTHTHT", "HTHHTHTT"),
	alphabet = 2)
 HTHTHTHT  HTHHTHTT 
0.2736772 0.5573217 
\end{verbatim}
\normalsize

\subsubsection{Entropy and second-order entropy}
Entropy \cite{barbasz2008estimate, shannon1948} has been used for decades as a measure of complexity. It must be emphasized that (first order) entropy does not capture the structure of a string, and only depends on the relative frequency of the symbols in the said string. For instance, the string ``0101010101010101'' has a greater entropy than ``0100101100100010'' because the first string is balanced in terms of 0's and 1's. According to entropy, the first string, although it is highly regular, should be considered more complex or more random than the second one.

Second-order entropy has been put forward to overcome this inconvenience, but it only does so partially. Indeed, second-order entropy only captures the narrowly local structure of a string. For instance, the string ``01100110011001100110...'' maximizes second-order entropy, because the four patterns 00, 01, 10 and 11 share the same frequency in this string. The fact that the sequence is the result of a simplistic rule is not taken into account. Notwithstanding these strong limitations, entropy has been extensively used. For that historical reason, \texttt{acss} includes two functions, \texttt{entropy()} and \texttt{entropy2()} (second order entropy).

\subsubsection{Change complexity}
Algorithmic complexity for short strings is an objective and universal normative measure of complexity approximating Kolmogorov-Chaitin complexity. ACSS helps in detecting \emph{any} computable departures from randomness. This is exactly what researchers seek when they want to assess the formal quality of a pseudo-random production. However, psychologists may also wish to assess complexity as it is perceived by human participants. In that case, algorithmic complexity may be too sensitive. For instance, there exists a (relatively) short program computing the decimals of $\pi$. However, faced with that series of digits, humans are unlikely to see any regularity: algorithmic complexity identifies as non-random some series that will look random to most humans.

When it comes to assessing \emph{perceived} complexity, the tool developed by Aksentijevic and Gibson \citeyear{aksentijevic2012}, named ``change complexity'', is an interesting alternative. It is based on the idea that humans' perception of complexity depends largely on the changes between one symbol and the next. Unfortunately, change complexity is, to date, only available for binary strings. As change complexity is to our knowledge not yet included in an \texttt{R}-package, we have implemented it in \texttt{acss} in function \texttt{change\_complexity()}.

\subsection{A comparison of complexity measures}
There are several complexity measures based on the coding theorem method, because the computation depends on the set of possible symbols the Turing machines can manipulate. To date, the package provides ACSS for 2, 4, 5, 6 and 9 symbols, giving 5 different measures of complexity. As we will see, however, these measures are strongly correlated. As a consequence one may use $K_{9}$ to assess the complexity of strings with 7 or 8 different symbols, or $K_{4}$ to assess the complexity of a string with 3 symbols. Thus in the end any alphabet size between 2 and 9 is available.  Also, these measures are mildly correlated to change complexity, and poorly to entropy.

\begin{figure}[t]
\begin{center}
\includegraphics[width=.5\textwidth]{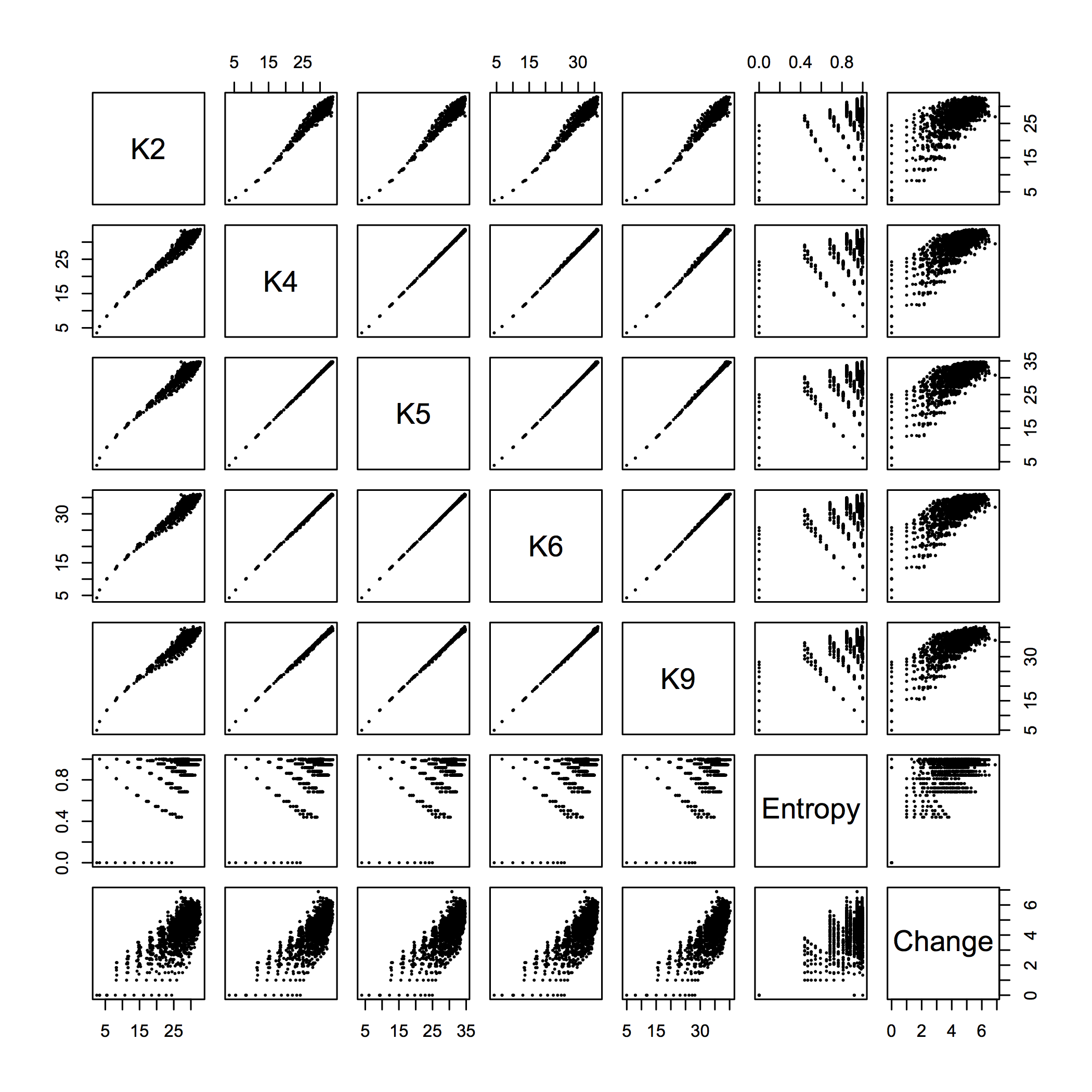}  
\caption{Scatter plot showing the relation between measures of complexity on every binary string with length from 1 to 11. ``Change'' stands for change complexity}
\label{fig:BinaryComp}
\end{center}
\end{figure}

Any binary string can be thought of as an $n$-symbol string ($n\ge 2$) that happens to use only 2 symbols. For instance, ``0101'' could be produced by a machine that only uses 0s and 1s, but also by a machine that uses digits from 0 to 9. Hence ``0101'' may be viewed as a word based on the alphabet $\{ 0,1\}$, but also based on $\{ 0,1, 2, 3, 4\}$, etc. Therefore, the complexity of a binary string can be rated by $K_{2}$, but also by $K_{4}$ or $K_{5}$. We computed $K_{n}$ (with $n\in \{ 2,4,5,6,9\}$), entropy and change complexity of all 2047 binary strings of length up to 11. Table \ref{tab:BinaryComp} displays the resulting correlations and Figure \ref{fig:BinaryComp} shows the corresponding scatter plots. The different algorithmic complexity estimates through the coding theorem method are closely related, with correlations above 0.999 for $K_{4},K_{5},K_{6}$ and $K_{9}$. $K_{2}$ is less correlated to the others, but every correlation stands above 0.97. There is a mild linear relation between ACSS and change complexity. Finally, entropy is only weakly linked to algorithmic and change complexity.

\begin{table}[b]
\centering
\begin{tabular}{lrrrrrr}
  \hline
 & K2 & K4 & K5 & K6 & K9 & Ent \\ 
  \hline
  K4 & 0.98 &  &  &  &  &    \\ 
  K5 & 0.97 & 1.00 &  &  &  &    \\ 
  K6 & 0.97 & 1.00 & 1.00 &  &  &    \\ 
  K9 & 0.97 & 1.00 & 1.00 & 1.00 &  &    \\ 
  Ent & 0.32 & 0.36 & 0.36 & 0.37 & 0.38 &  \\ 
  Change & 0.69 & 0.75 & 0.75 & 0.75 & 0.75 & 0.50 \\ 
   \hline
\end{tabular}
\caption{Correlation matrix of complexity measures computed on all binary strings of length up to 11. ``Ent'' stands for entropy, and ``Change'' for change complexity.}
\label{tab:BinaryComp}
\end{table}

\begin{figure}[t]
\begin{center}
\includegraphics[width=0.5\textwidth]{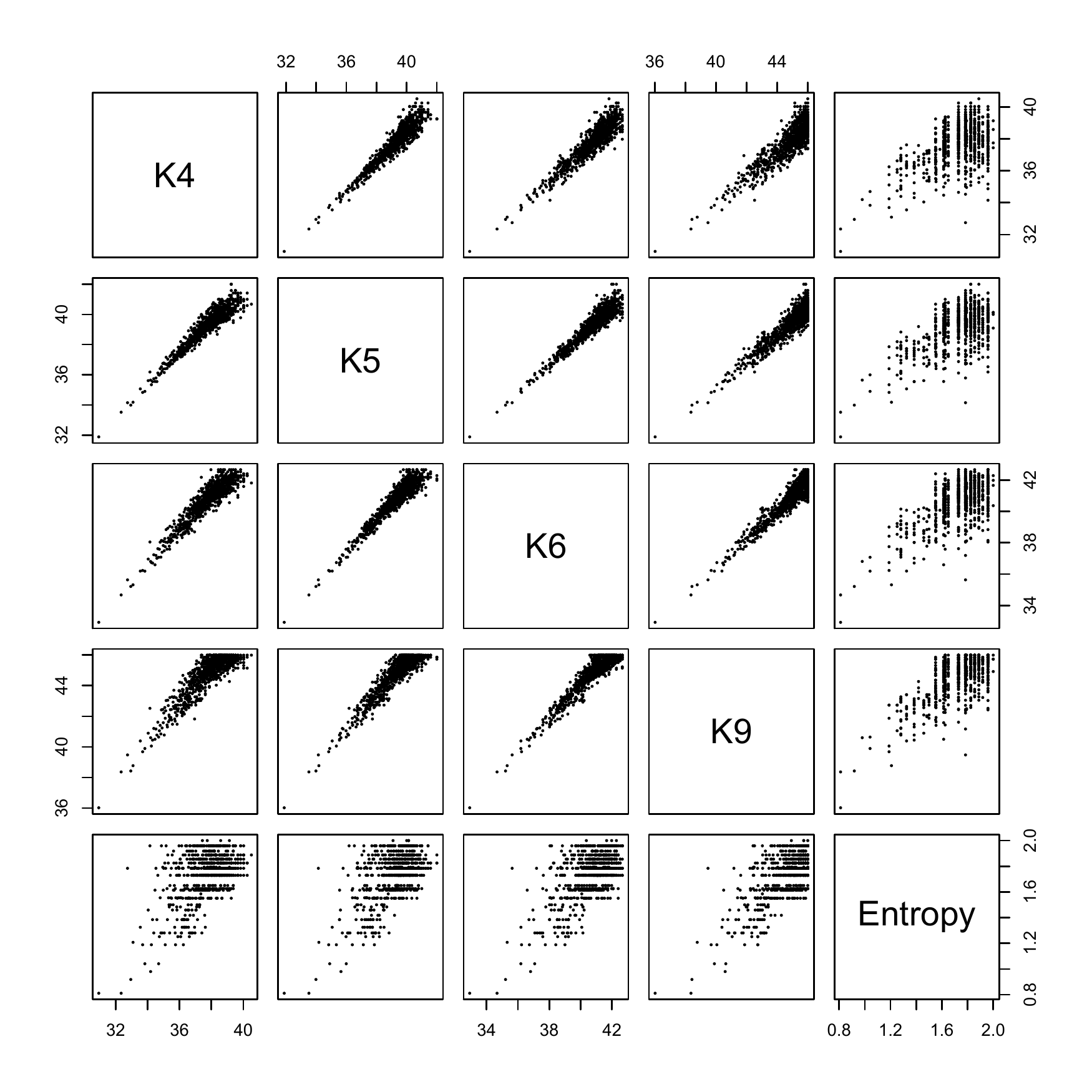}
\caption{Scatterplot matrix computed with 844 randomly chosen 12-character long 4-symbol strings.}
\label{fig:FourSymbolsComp}
\end{center}
\end{figure}

The correlation between different versions of ACSS (that is, $K_{n}$) may be partly explained by the length of the strings. Unlike entropy, ACSS is sensitive to the number of symbols in a string. This is not a weakness. On the contrary, if the complexity of a string is linked to the evidence it brings to the chance hypothesis, it should depend on length. Throwing a coin to determine if it is fair and getting HTHTTTHH is more convincing than getting just HT. Although both strings are balanced, the first one should be considered more complex because it affords more evidence that the coin is fair (and not, for instance, bound to alternate heads and tails). However, to control for the effect of length and extend our previous result to more complex strings, we picked up 844 random 4-symbol strings of length 12. Change complexity is not defined for non-binary sequences, but as Figure \ref{fig:FourSymbolsComp}  and Table \ref{tab:FourSymbolsComp} illustrate, the different ACSS measures are strongly correlated, and mildly correlated to entropy.

\begin{table}[b]
\centering
\begin{tabular}{lrrrr}
  \hline
 & K4 & K5 & K6 & K9  \\ 
  \hline 
  K5 & 0.94 &  &  &   \\ 
  K6 & 0.92 & 0.95 &  &   \\ 
  K9 & 0.88 & 0.92 & 0.94 &   \\ 
  Entropy & 0.52 & 0.58 & 0.62 & 0.69  \\ 
   \hline
\end{tabular}
\caption{Pearson's correlation between complexity measures, computed from 844 random 12-character long 4-symbol strings.}
\label{tab:FourSymbolsComp}   
\end{table}

\section{Applications}
In this last section, we provide three illustrative applications of ACSS. The first two are short reports of new and illustrative experiments and the third one is a re-analysis of previously published data \cite{matthews2013}. Although these experiments are presented to illustrate the use of ACSS, they also provide new insights into subjective probability and the perception of randomness. Note that all the data and analysis scripts for these applications are also part of \texttt{acss}.

\subsection{Experiment 1: Humans are ``better than chance''}
Human pseudo-random sequential binary productions have been reported to be overly complex, in the sense that they are more complex than the average complexity of truly random sequences \cite<i.e., sequences of fixed length produced by repeatedly tossing a coin;>{gauvrit2014}. Here, we test the same effect with non-binary sequences based on 4 symbols. To replicate the analysis, type \texttt{?exp1} at the \texttt{R} prompt after loading \texttt{acss} and execute the examples.
\subsubsection{Participants}
A sample of 34 healthy adults participated in this experiment. Ages ranged from 20 to 55 (mean $= 37.65$, SD $= 7.98$). Participants were recruited via e-mail and did not receive any compensation for their participation.
\subsubsection{Methods}
Participants were asked to produce at their own pace a series of 10 symbols using ``A'', ``B'', ``C'', and ``D''  that would ``look as random as possible, so that if someone else saw the sequence, she would believe it to be a truly random one''. Participants submitted their responses via e-mail.
\subsubsection{Results}
\begin{figure}[t]
\begin{center}
\includegraphics[width=0.5\textwidth]{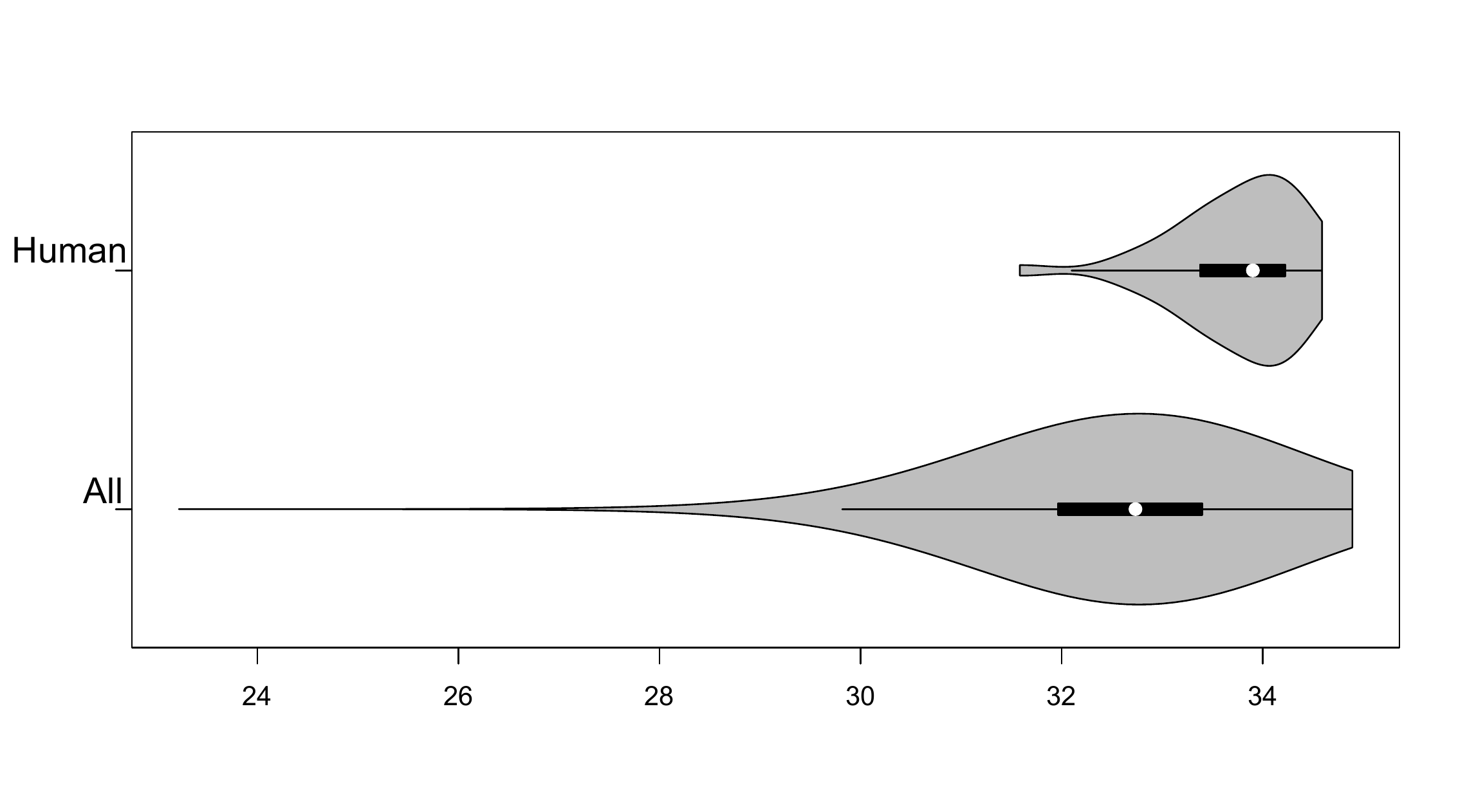}
\caption{Violin plot showing the distribution of complexity of human strings vs. every possible pattern of strings, with 4-symbol alphabet and length 10.}
\label{fig:violinplot}
\end{center}
\end{figure}

A one sample $t$-test showed that the mean complexity of the responses of participants is significantly larger that the mean complexity of all possible patterns of length 10 ($t(33) = 10.62, p < .0001$). The violin plot in Figure \ref{fig:violinplot} shows that human productions are more complex than random patterns because humans avoid low-complexity strings. On the other hand, human productions did not reach the highest possible values of complexity.

\subsubsection{Discussion}
These results are consistent with the hypothesis that when participants try to behave randomly, they in fact tend to maximize the complexity of their responses, leading to overly complex sequences. However, whereas they succeed in avoiding low-complexity patterns, they cannot build the most complex strings.

\subsection{Experiment 2: The threshold of complexity -- a case study}
Humans are sensitive to regularity and distinguish truly random series from deterministic ones \cite{yamada_pattern_2013}. More complex strings should be more likely to be considered random than simple ones. Here, we briefly address this question through a binary forced choice task. We assume that there exists an individual threshold of complexity for which the probability that the individual identifies a string as random is $.5$. We estimated that threshold for one participant. The participant was a healthy adult male, 42 years old. The data and code are available by calling \texttt{?exp2}.

\subsubsection{Methods}
A series of 200 random strings of length 10 from an alphabet of 6 symbols, such as ``6154256554'', were generated with the \texttt{R} function \texttt{sample()}. For each string, the participant had to decide whether or not the sequence appeared random.

\subsubsection{Results}
\begin{figure}[b]
\begin{center}
\includegraphics[width=0.5\textwidth]{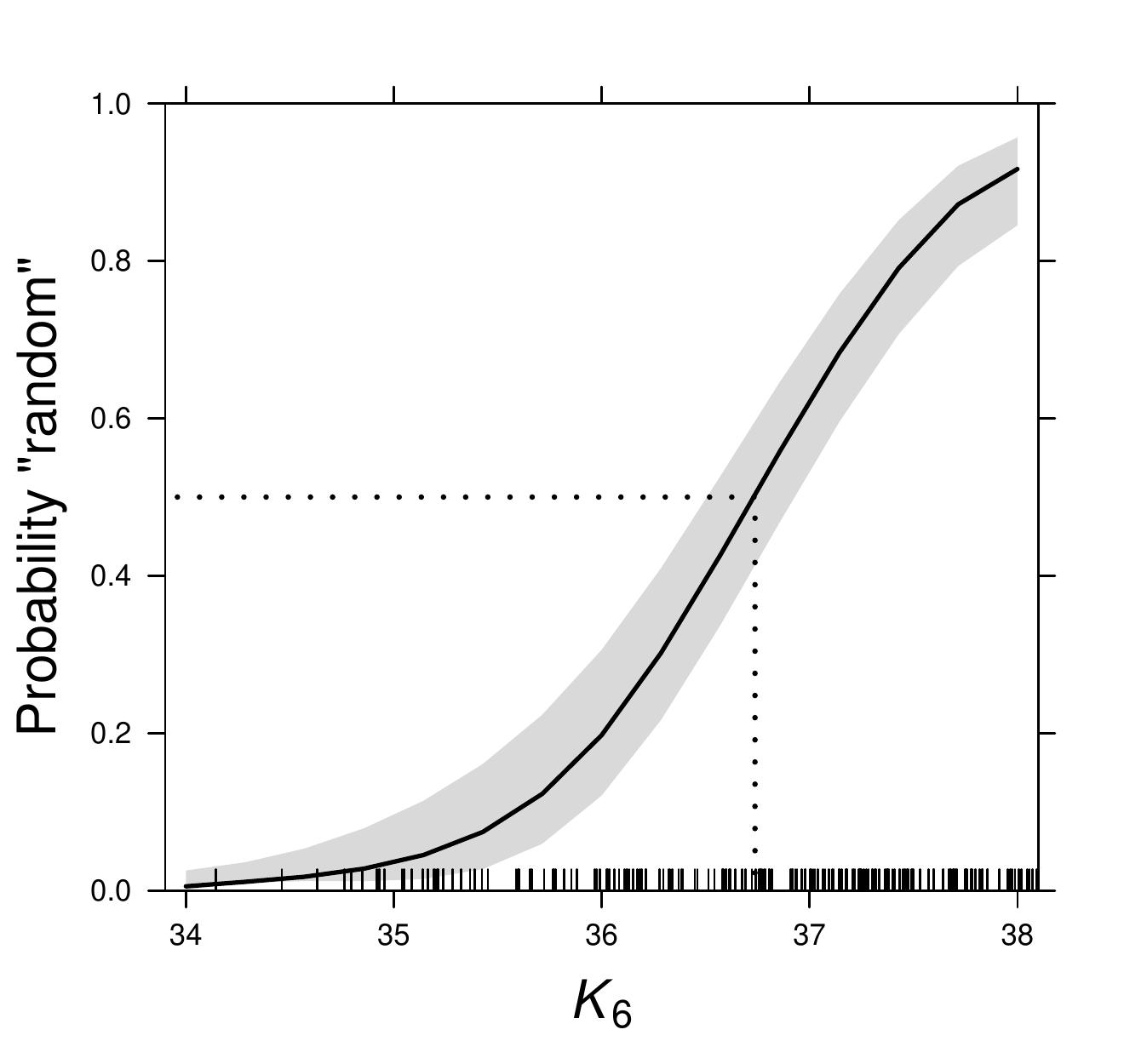}
\caption{Graphical display of the logistic regression with actual complexities ($K_6$) of 200 strings as independent variable and the observed responses (appears random or not) of one participant as dependent variable. The gray area depicts 95\%-confidence bands, the black dots at the bottom the 200 complexities. The dotted lines show the threshold where the perceived probability of randomness is 0.5.}
\label{fig:threshold}
\end{center}
\end{figure}

A logistic regression of the actual complexities of the strings ($K_6$) on the responses is displayed in Figure~\ref{fig:threshold}. The results showed that more complex sequences were more likely to be considered random (slope $=1.9$, $p< .0001$, corresponding to an odds ratio of 6.69). Furthermore, a complexity of 36.74 corresponded to the subjective probability of randomness 0.5 (i.e., the individual threshold was 36.74). 

\subsection{The span of local complexity}
In a study of contextual effect in the perception of randomness, Matthews (\citeyearNP{matthews2013}, Experiment 1) showed participants series of binary strings of length 21. For each string, participants had to rate the sequence on a 6-point scale ranging from ``definitely random'' to ``definitely not random''. Results showed that participants were influenced by the context of presentation: sequences with medial alternation rate (AR) were considered highly random when they were intermixed with low AR, but as relatively non-random when intermixed with high AR sequences. In the following, we will analyze the data irrespective of the context or AR.

When individuals judge whether a short string of, for example, 3-6 characters, is random, they probably consider the complete sequence. For these cases, ACSS would be the right normative measure. When strings are longer, such as a length of 21, individuals probably cannot consider the complete sequence at once. Matthews (2013) and others \cite<e.g.,>{hahn_experiential_2014} have hypothesized that in these cases individuals rely on the local complexity of the string. If this were true, the question remains as to how local the analysis is. To answer this, we will reanalyze Matthews' data.

For each string and each span ranging from 3 to 11, we first computed the mean local complexity of the string. For instance, the string ``XXXXXXXOOOOXXXOOOOOOO'' with span 11 gives a mean local complexity of 29.53. The same string has a mean local complexity of 11.22 with span 5.

\footnotesize
\begin{verbatim}
> sapply(local_complexity("XXXXXXXOOOOXXXOOOOOOO",
	11, 2), mean)
XXXXXXXOOOOXXXOOOOOOO 
             29.52912 
> sapply(local_complexity("XXXXXXXOOOOXXXOOOOOOO",
	5, 2), mean)
XXXXXXXOOOOXXXOOOOOOO 
             11.21859 
\end{verbatim}
\normalsize

\begin{figure}[htbp]
\begin{center}
\includegraphics[width=0.5\textwidth]{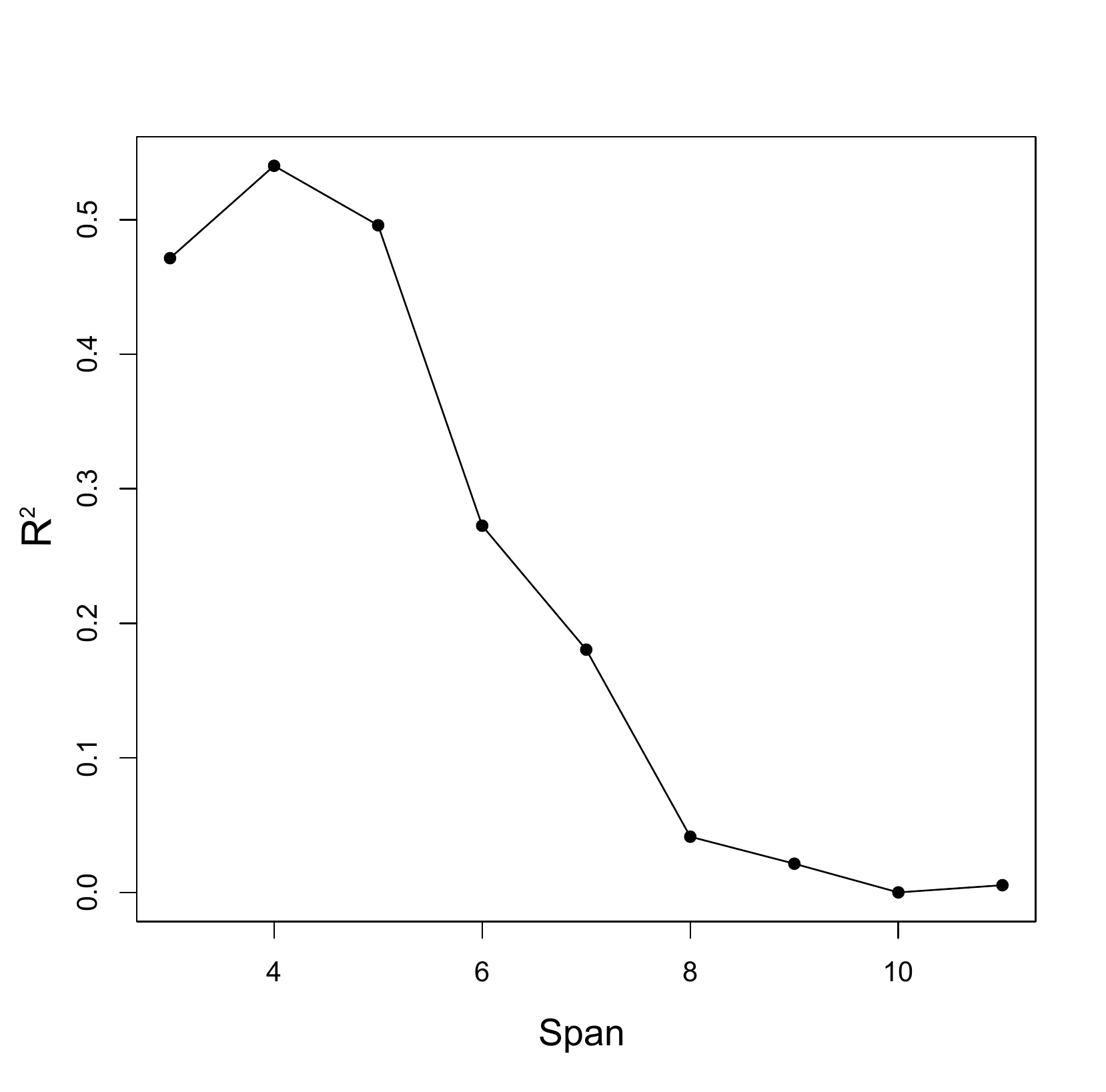}
\caption{$R^{2}$ between mean local complexity with span 3 to 11 and the subjective mean evaluation of randomness.}
\label{fig:DeterminationBySpan}
\end{center}
\end{figure}

For each span, we then computed $R^{2}$ (the proportion of variance accounted for) between mean local complexity (a formal measure) and the mean randomness score given by the participants in Matthews' (2013) Experiment 1. Figure~\ref{fig:DeterminationBySpan} shows that a span of 4 or 5 best describes the judgments with $R^{2}$ of 54\% and 50\%. Furthermore, $R^{2}$ decreases so fast that it amounts to less than 0.002\% when the span is set to 10. These results suggest that when asked to judge if a string is random, individuals rely on very local structural features of the strings, only considering subsequences of 4-5 symbols. This is very near the suggested limit of the short term memory of 4 chunks (Cowan, 2001). Future researchers could build on this preliminary account to investigate the possible ``span'' of human observation in the face of possibly random serial data. The data and code for this application are available by calling \texttt{?matthews2013}.

\section{Relationship to complexity based model selection}

As mentioned in the beginning, complexity also plays an important role in modern approaches to model selection, specifically within the minimum description length framework~\cite<MDL;>{grunwald2007minimum, myung_model_2006}. Model selection in this context~\cite<see e.g.>{myung2014} refers to the process of selecting, among a set of candidate models, the model that strikes the best balance between goodness-of-fit (i.e., how well does the model describe the obtained data) and complexity (i.e., how well does the model describe all possible data or data in general). MDL provides a principled way of combining model fit with a quantification of model complexity that originated in information theory~\cite{rissanen1989stochastic} and is, similar to algorithmic complexity, based on the notion of compressed code. The basic idea is that a model can be viewed as an algorithm that, in combination with a set of parameters, can produce a specific prediction. A model that describes the observed data well (i.e., provides a good fit) is a model that can compress the data well as it only requires a set of parameters and there are little residuals that need to be described in addition. Within this framework, the complexity of a model is the shortest possible code or algorithm that describes all possible data pattern predicted by the model. The model selection index is the length of the concatenation of the code describing parameters and residuals and the code producing all possible data sets. As usual, the best model in terms of MDL is the model with the lowest model selection index. 

The MDL approach differs from the complexity approach discussed in the current manuscript as it focuses on a specific set of models. To be selected as possible candidates, models usually need to satisfy other criteria such as providing explanatory value, need to be a priori plausible, and the parameters should be interpretable in terms of psychological processes \cite{myung2014}. In contrast, algorithmic complexity is concerned with finding the shortest possible description considering all possible models leaving those considerations aside. However, given that both approaches are couched within the same information theoretic framework, MDL converges towards algorithmic complexity if the set of candidate models becomes infinite~\cite{wallace1999minimum}. Furthermore, in the current manuscript we are only concerned with short strings, whereas even in compressed form data and the prediction space of a model are usually comparatively large.

\section{Conclusion}
Until the development of the coding theorem method \cite{soler2012,delahaye2012}, researchers interested in short random strings were constrained to use measures of complexity that focused on particular features of randomness. This has led to a number of (unsatisfactory) measures \cite{towse1998}. Each of these previously used measures can be thought of as a way of performing a particular statistical test of randomness. In contrast, algorithmic complexity affords access to the ultimate measure of randomness, as the algorithmic complexity definition of randomness has been shown to be equivalent to defining a random sequence as one which would pass \emph{every} computable test of randomness. We have computed an approximation of algorithmic complexity for short strings and made this approximation freely available in the \texttt{R} package \texttt{acss}.

Because human capacities are limited, it is unlikely that humans will be able to recognize every kind of deviation from randomness. Subjective randomness does not equal algorithmic complexity, as Griffiths and Tenenbaum \citeyear{griffiths2004} remind us. Other measures of complexity will still be useful in describing the ways in which human pseudo-random behaviors or the subjective perception of randomness or complexity differ from objective randomness, as defined within the mathematical theory of randomness and advocated in this manuscript. But to achieve this goal of comparing subjective and objective randomness, we also need an objective and universal measure of randomness (or complexity) based on a sound mathematical theory of randomness. Although the uncomputability of Kolmogorov complexity places some limitations on what is knowable about objective randomness, ACSS provides a sensible and practical approximation that researchers can use in real life. We are confident that ACSS will prove very useful as a normative measure of complexity in helping psychologists understand how human subjective randomness differs from  ``true'' randomness as defined by algorithmic complexity.

\bibliography{ComplexityForPsychologyBib}

\begin{thebibliography}{}

\bibitem[\protect\citeauthoryear{%
Aksentijevic%
\ \BBA{} Gibson%
}{%
Aksentijevic%
\ \BBA{} Gibson%
}{%
{\protect\APACyear{2012}}%
}]{%
aksentijevic2012}%
\APACinsertmetastar{%
aksentijevic2012}%
Aksentijevic, A.%
\BCBT{}\ \BBA{} Gibson, K.%
%
\unskip\
\newblock
\APACrefYearMonthDay{2012}{}{}.
\newblock
\BBOQ{}\APACrefatitle{Complexity equals change.}{Complexity equals
  change.}\BBCQ{}
\newblock
\APACjournalVolNumPages{Cognitive Systems Research}{15-16}{}{1--16}.
\PrintBackRefs{\CurrentBib}

\bibitem[\protect\citeauthoryear{%
Audiffren%
, Tomporowski%
\BCBL{}\ \BBA{} Zagrodnik%
}{%
Audiffren%
\ \protect\BOthers{.}}{%
{\protect\APACyear{2009}}%
}]{%
audiffren2009}%
\APACinsertmetastar{%
audiffren2009}%
Audiffren, M.%
, Tomporowski, P\BPBI D.%
\BCBL{}\ \BBA{} Zagrodnik, J.%
%
\unskip\
\newblock
\APACrefYearMonthDay{2009}{}{}.
\newblock
\BBOQ{}\APACrefatitle{Acute aerobic exercise and information processing:
  modulation of executive control in a random number generation task}{Acute
  aerobic exercise and information processing: modulation of executive control
  in a random number generation task}.\BBCQ{}
\newblock
\APACjournalVolNumPages{Acta Psychologica}{132}{1}{85--95}.
\PrintBackRefs{\CurrentBib}

\bibitem[\protect\citeauthoryear{%
Baddeley%
, Thomson%
\BCBL{}\ \BBA{} Buchanan%
}{%
Baddeley%
\ \protect\BOthers{.}}{%
{\protect\APACyear{1975}}%
}]{%
baddeley1975word}%
\APACinsertmetastar{%
baddeley1975word}%
Baddeley, A\BPBI D.%
, Thomson, N.%
\BCBL{}\ \BBA{} Buchanan, M.%
%
\unskip\
\newblock
\APACrefYearMonthDay{1975}{}{}.
\newblock
\BBOQ{}\APACrefatitle{Word length and the structure of short-term memory}{Word
  length and the structure of short-term memory}.\BBCQ{}
\newblock
\APACjournalVolNumPages{Journal of Verbal Learning and Verbal
  Behavior}{14}{6}{575--589}.
\PrintBackRefs{\CurrentBib}

\bibitem[\protect\citeauthoryear{%
Barbasz%
, Stettner%
, Wierzcho{\'n}%
, Piotrowski%
\BCBL{}\ \BBA{} Barbasz%
}{%
Barbasz%
\ \protect\BOthers{.}}{%
{\protect\APACyear{2008}}%
}]{%
barbasz2008estimate}%
\APACinsertmetastar{%
barbasz2008estimate}%
Barbasz, J.%
, Stettner, Z.%
, Wierzcho{\'n}, M.%
, Piotrowski, K\BPBI T.%
\BCBL{}\ \BBA{} Barbasz, A.%
%
\unskip\
\newblock
\APACrefYearMonthDay{2008}{}{}.
\newblock
\BBOQ{}\APACrefatitle{How to estimate the randomness in random sequence
  generation tasks?.}{How to estimate the randomness in random sequence
  generation tasks?.}\BBCQ{}
\newblock
\APACjournalVolNumPages{Polish Psychological Bulletin}{39}{1}{42 - 46}.
\PrintBackRefs{\CurrentBib}

\bibitem[\protect\citeauthoryear{%
B{\'e}dard%
, Joyal%
, Godbout%
\BCBL{}\ \BBA{} Chantal%
}{%
B{\'e}dard%
\ \protect\BOthers{.}}{%
{\protect\APACyear{2009}}%
}]{%
bedard2009}%
\APACinsertmetastar{%
bedard2009}%
B{\'e}dard, M\BHBI J.%
, Joyal, C\BPBI C.%
, Godbout, L.%
\BCBL{}\ \BBA{} Chantal, S.%
%
\unskip\
\newblock
\APACrefYearMonthDay{2009}{}{}.
\newblock
\BBOQ{}\APACrefatitle{Executive functions and the obsessive-compulsive
  disorder: On the importance of subclinical symptoms and other concomitant
  factors}{Executive functions and the obsessive-compulsive disorder: On the
  importance of subclinical symptoms and other concomitant factors}.\BBCQ{}
\newblock
\APACjournalVolNumPages{Archives of Clinical Neuropsychology}{24}{6}{585--598}.
\PrintBackRefs{\CurrentBib}

\bibitem[\protect\citeauthoryear{%
Bianchi%
\ \BBA{} Mendez%
}{%
Bianchi%
\ \BBA{} Mendez%
}{%
{\protect\APACyear{2013}}%
}]{%
bianchi2013}%
\APACinsertmetastar{%
bianchi2013}%
Bianchi, A\BPBI M.%
\BCBT{}\ \BBA{} Mendez, M\BPBI O.%
%
\unskip\
\newblock
\APACrefYearMonthDay{2013}{}{}.
\newblock
\BBOQ{}\APACrefatitle{Methods for heart rate variability analysis during
  sleep}{Methods for heart rate variability analysis during sleep}.\BBCQ{}
\newblock
\BIn{} \APACrefbtitle{Engineering in Medicine and Biology Society (EMBC), 2013
  35th Annual International Conference of the IEEE}{Engineering in medicine and
  biology society (embc), 2013 35th annual international conference of the
  ieee}\ (\BPGS\ 6579--6582).
\PrintBackRefs{\CurrentBib}

\bibitem[\protect\citeauthoryear{%
Boon%
, Casti%
\BCBL{}\ \BBA{} Taylor%
}{%
Boon%
\ \protect\BOthers{.}}{%
{\protect\APACyear{2011}}%
}]{%
boon2011}%
\APACinsertmetastar{%
boon2011}%
Boon, J\BPBI P.%
, Casti, J.%
\BCBL{}\ \BBA{} Taylor, R\BPBI P.%
%
\unskip\
\newblock
\APACrefYearMonthDay{2011}{}{}.
\newblock
\BBOQ{}\APACrefatitle{Artistic forms and complexity}{Artistic forms and
  complexity}.\BBCQ{}
\newblock
\APACjournalVolNumPages{Nonlinear Dynamics-Psychology and Life
  Sciences}{15}{2}{265}.
\PrintBackRefs{\CurrentBib}

\bibitem[\protect\citeauthoryear{%
Brandouy%
, Delahaye%
, Ma%
\BCBL{}\ \BBA{} Zenil%
}{%
Brandouy%
\ \protect\BOthers{.}}{%
{\protect\APACyear{2012}}%
}]{%
brandouy2012}%
\APACinsertmetastar{%
brandouy2012}%
Brandouy, O.%
, Delahaye, J\BHBI P.%
, Ma, L.%
\BCBL{}\ \BBA{} Zenil, H.%
%
\unskip\
\newblock
\APACrefYearMonthDay{2012}{}{}.
\newblock
\BBOQ{}\APACrefatitle{Algorithmic complexity of financial motions}{Algorithmic
  complexity of financial motions}.\BBCQ{}
\newblock
\APACjournalVolNumPages{Research in International Business and
  Finance}{30}{C}{336--347}.
\PrintBackRefs{\CurrentBib}

\bibitem[\protect\citeauthoryear{%
Brown%
\ \BBA{} Marsden%
}{%
Brown%
\ \BBA{} Marsden%
}{%
{\protect\APACyear{1990}}%
}]{%
brown1990}%
\APACinsertmetastar{%
brown1990}%
Brown, R.%
\BCBT{}\ \BBA{} Marsden, C.%
%
\unskip\
\newblock
\APACrefYearMonthDay{1990}{}{}.
\newblock
\BBOQ{}\APACrefatitle{Cognitive function in Parkinson's disease: from
  description to theory}{Cognitive function in parkinson's disease: from
  description to theory}.\BBCQ{}
\newblock
\APACjournalVolNumPages{Trends in Neurosciences}{13}{1}{21--29}.
\PrintBackRefs{\CurrentBib}

\bibitem[\protect\citeauthoryear{%
Calude%
}{%
Calude%
}{%
{\protect\APACyear{2002}}%
}]{%
calude2002}%
\APACinsertmetastar{%
calude2002}%
Calude, C.%
%
\unskip\
\newblock
\APACrefYear{2002}.
\newblock
\APACrefbtitle{Information and Randomness. An Algorithmic
  Perspective}{Information and randomness. an algorithmic perspective}\
  (\PrintOrdinal{2nd, revised and extended}\ \BEd).
\newblock
\APACaddressPublisher{}{Springer-Verlag}.
\PrintBackRefs{\CurrentBib}

\bibitem[\protect\citeauthoryear{%
Cardaci%
, Di~Gesu%
, Petrou%
\BCBL{}\ \BBA{} Tabacchi%
}{%
Cardaci%
\ \protect\BOthers{.}}{%
{\protect\APACyear{2009}}%
}]{%
cardaci2009}%
\APACinsertmetastar{%
cardaci2009}%
Cardaci, M.%
, Di~Gesu, V.%
, Petrou, M.%
\BCBL{}\ \BBA{} Tabacchi, M\BPBI E.%
%
\unskip\
\newblock
\APACrefYearMonthDay{2009}{}{}.
\newblock
\BBOQ{}\APACrefatitle{Attentional vs computational complexity measures in
  observing paintings}{Attentional vs computational complexity measures in
  observing paintings}.\BBCQ{}
\newblock
\APACjournalVolNumPages{Spatial vision}{22}{3}{195--209}.
\PrintBackRefs{\CurrentBib}

\bibitem[\protect\citeauthoryear{%
Chaitin%
}{%
Chaitin%
}{%
{\protect\APACyear{1966}}%
}]{%
chaitin1966}%
\APACinsertmetastar{%
chaitin1966}%
Chaitin, G.%
%
\unskip\
\newblock
\APACrefYearMonthDay{1966}{}{}.
\newblock
\BBOQ{}\APACrefatitle{On the length of programs for computing finite binary
  sequences}{On the length of programs for computing finite binary
  sequences}.\BBCQ{}
\newblock
\APACjournalVolNumPages{Journal of the ACM}{13}{4}{547-569}.
\PrintBackRefs{\CurrentBib}

\bibitem[\protect\citeauthoryear{%
Chaitin%
}{%
Chaitin%
}{%
{\protect\APACyear{2004}}%
}]{%
chaitin2004}%
\APACinsertmetastar{%
chaitin2004}%
Chaitin, G.%
%
\unskip\
\newblock
\APACrefYear{2004}.
\newblock
\APACrefbtitle{Algorithmic information theory}{Algorithmic information theory}\
  (\BVOL~1).
\newblock
\APACaddressPublisher{}{Cambridge University Press}.
\PrintBackRefs{\CurrentBib}

\bibitem[\protect\citeauthoryear{%
Chater%
}{%
Chater%
}{%
{\protect\APACyear{1996}}%
}]{%
chater1996}%
\APACinsertmetastar{%
chater1996}%
Chater, N.%
%
\unskip\
\newblock
\APACrefYearMonthDay{1996}{}{}.
\newblock
\BBOQ{}\APACrefatitle{Reconciling simplicity and likelihood principles in
  perceptual organization}{Reconciling simplicity and likelihood principles in
  perceptual organization}.\BBCQ{}
\newblock
\APACjournalVolNumPages{Psychological Review}{103}{3}{566-581}.
\PrintBackRefs{\CurrentBib}

\bibitem[\protect\citeauthoryear{%
Chater%
\ \BBA{} Vit\'{a}nyi%
}{%
Chater%
\ \BBA{} Vit\'{a}nyi%
}{%
{\protect\APACyear{2003}}%
}]{%
chater2003}%
\APACinsertmetastar{%
chater2003}%
Chater, N.%
\BCBT{}\ \BBA{} Vit\'{a}nyi, P.%
%
\unskip\
\newblock
\APACrefYearMonthDay{2003}{}{}.
\newblock
\BBOQ{}\APACrefatitle{Simplicity: A unifying principle in cognitive
  science?}{Simplicity: A unifying principle in cognitive science?}\BBCQ{}
\newblock
\APACjournalVolNumPages{Trends in Cognitive Sciences}{7}{1}{19-22}.
\PrintBackRefs{\CurrentBib}

\bibitem[\protect\citeauthoryear{%
Cilibrasi%
\ \BBA{} Vit{\'a}nyi%
}{%
Cilibrasi%
\ \BBA{} Vit{\'a}nyi%
}{%
{\protect\APACyear{2005}}%
}]{%
cilibrasi2005}%
\APACinsertmetastar{%
cilibrasi2005}%
Cilibrasi, R.%
\BCBT{}\ \BBA{} Vit{\'a}nyi, P.%
%
\unskip\
\newblock
\APACrefYearMonthDay{2005}{}{}.
\newblock
\BBOQ{}\APACrefatitle{Clustering by compression}{Clustering by
  compression}.\BBCQ{}
\newblock
\APACjournalVolNumPages{Information Theory, {IEEE} Transactions
  on}{51}{4}{1523--1545}.
\PrintBackRefs{\CurrentBib}

\bibitem[\protect\citeauthoryear{%
Cilibrasi%
\ \BBA{} Vit{\'a}nyi%
}{%
Cilibrasi%
\ \BBA{} Vit{\'a}nyi%
}{%
{\protect\APACyear{2007}}%
}]{%
cilibrasi2007}%
\APACinsertmetastar{%
cilibrasi2007}%
Cilibrasi, R.%
\BCBT{}\ \BBA{} Vit{\'a}nyi, P.%
%
\unskip\
\newblock
\APACrefYearMonthDay{2007}{}{}.
\newblock
\BBOQ{}\APACrefatitle{The google similarity distance}{The google similarity
  distance}.\BBCQ{}
\newblock
\APACjournalVolNumPages{Knowledge and Data Engineering, IEEE Transactions
  on}{19}{3}{370--383}.
\PrintBackRefs{\CurrentBib}

\bibitem[\protect\citeauthoryear{%
Cowan%
}{%
Cowan%
}{%
{\protect\APACyear{2001}}%
}]{%
cowan2001magical}%
\APACinsertmetastar{%
cowan2001magical}%
Cowan, N.%
%
\unskip\
\newblock
\APACrefYearMonthDay{2001}{}{}.
\newblock
\BBOQ{}\APACrefatitle{The magical number 4 in short-term memory: A
  reconsideration of mental storage capacity}{The magical number 4 in
  short-term memory: A reconsideration of mental storage capacity}.\BBCQ{}
\newblock
\APACjournalVolNumPages{Behavioral and Brain Sciences}{24}{1}{87--114}.
\PrintBackRefs{\CurrentBib}

\bibitem[\protect\citeauthoryear{%
Crova%
\ \protect\BOthers{.}}{%
Crova%
\ \protect\BOthers{.}}{%
{\protect\APACyear{2013}}%
}]{%
crova2013}%
\APACinsertmetastar{%
crova2013}%
Crova, C.%
, Struzzolino, I.%
, Marchetti, R.%
, Masci, I.%
, Vannozzi, G.%
, Forte, R.%
\BCBL{}\ \BOthersPeriod{.}%
\unskip\
\newblock
\APACrefYearMonthDay{2013}{}{}.
\newblock
\BBOQ{}\APACrefatitle{Cognitively challenging physical activity benefits
  executive function in overweight children}{Cognitively challenging physical
  activity benefits executive function in overweight children}.\BBCQ{}
\newblock
\APACjournalVolNumPages{Journal of Sports Sciences}{ahead-of-print}{}{1--11}.
\PrintBackRefs{\CurrentBib}

\bibitem[\protect\citeauthoryear{%
Curci%
, Lanciano%
, Soleti%
\BCBL{}\ \BBA{} Rim{\'e}%
}{%
Curci%
\ \protect\BOthers{.}}{%
{\protect\APACyear{2013}}%
}]{%
curci2013}%
\APACinsertmetastar{%
curci2013}%
Curci, A.%
, Lanciano, T.%
, Soleti, E.%
\BCBL{}\ \BBA{} Rim{\'e}, B.%
%
\unskip\
\newblock
\APACrefYearMonthDay{2013}{}{}.
\newblock
\BBOQ{}\APACrefatitle{Negative Emotional Experiences Arouse Rumination and
  Affect Working Memory Capacity}{Negative emotional experiences arouse
  rumination and affect working memory capacity}.\BBCQ{}
\newblock
\APACjournalVolNumPages{Emotion}{13}{5}{867--880}.
\PrintBackRefs{\CurrentBib}

\bibitem[\protect\citeauthoryear{%
Delahaye%
\ \BBA{} Zenil%
}{%
Delahaye%
\ \BBA{} Zenil%
}{%
{\protect\APACyear{2012}}%
}]{%
delahaye2012}%
\APACinsertmetastar{%
delahaye2012}%
Delahaye, J\BHBI P.%
\BCBT{}\ \BBA{} Zenil, H.%
%
\unskip\
\newblock
\APACrefYearMonthDay{2012}{}{}.
\newblock
\BBOQ{}\APACrefatitle{Numerical evaluation of algorithmic complexity for short
  strings: A glance into the innermost structure of randomness}{Numerical
  evaluation of algorithmic complexity for short strings: A glance into the
  innermost structure of randomness}.\BBCQ{}
\newblock
\APACjournalVolNumPages{Applied Mathematics and Computation}{219}{1}{63--77}.
\PrintBackRefs{\CurrentBib}

\bibitem[\protect\citeauthoryear{%
Downey%
\ \BBA{} Hirschfeldt%
}{%
Downey%
\ \BBA{} Hirschfeldt%
}{%
{\protect\APACyear{2008}}%
}]{%
downey2008}%
\APACinsertmetastar{%
downey2008}%
Downey, R\BPBI R\BPBI G.%
\BCBT{}\ \BBA{} Hirschfeldt, D\BPBI R.%
%
\unskip\
\newblock
\APACrefYear{2008}.
\newblock
\APACrefbtitle{Algorithmic randomness and complexity}{Algorithmic randomness
  and complexity}.
\newblock
\APACaddressPublisher{}{Springer}.
\PrintBackRefs{\CurrentBib}

\bibitem[\protect\citeauthoryear{%
Elzinga%
}{%
Elzinga%
}{%
{\protect\APACyear{2010}}%
}]{%
elzinga2010}%
\APACinsertmetastar{%
elzinga2010}%
Elzinga, C\BPBI H.%
%
\unskip\
\newblock
\APACrefYearMonthDay{2010}{}{}.
\newblock
\BBOQ{}\APACrefatitle{Complexity of categorical time series}{Complexity of
  categorical time series}.\BBCQ{}
\newblock
\APACjournalVolNumPages{Sociological Methods {\&} Research}{38}{3}{463--481}.
\PrintBackRefs{\CurrentBib}

\bibitem[\protect\citeauthoryear{%
Feldman%
}{%
Feldman%
}{%
{\protect\APACyear{2000}}%
}]{%
feldman2000}%
\APACinsertmetastar{%
feldman2000}%
Feldman, J.%
%
\unskip\
\newblock
\APACrefYearMonthDay{2000}{}{}.
\newblock
\BBOQ{}\APACrefatitle{Minimization of Boolean complexity in human concept
  learning}{Minimization of boolean complexity in human concept
  learning}.\BBCQ{}
\newblock
\APACjournalVolNumPages{Nature}{407}{6804}{630--633}.
\PrintBackRefs{\CurrentBib}

\bibitem[\protect\citeauthoryear{%
Feldman%
}{%
Feldman%
}{%
{\protect\APACyear{2003}}%
}]{%
feldman2003}%
\APACinsertmetastar{%
feldman2003}%
Feldman, J.%
%
\unskip\
\newblock
\APACrefYearMonthDay{2003}{}{}.
\newblock
\BBOQ{}\APACrefatitle{A catalog of Boolean concepts}{A catalog of boolean
  concepts}.\BBCQ{}
\newblock
\APACjournalVolNumPages{Journal of Mathematical Psychology}{47}{1}{75-89}.
\PrintBackRefs{\CurrentBib}

\bibitem[\protect\citeauthoryear{%
Feldman%
}{%
Feldman%
}{%
{\protect\APACyear{2006}}%
}]{%
feldman2006}%
\APACinsertmetastar{%
feldman2006}%
Feldman, J.%
%
\unskip\
\newblock
\APACrefYearMonthDay{2006}{}{}.
\newblock
\BBOQ{}\APACrefatitle{An algebra of human concept learning}{An algebra of human
  concept learning}.\BBCQ{}
\newblock
\APACjournalVolNumPages{Journal of Mathematical Psychology}{50}{4}{339-368}.
\PrintBackRefs{\CurrentBib}

\bibitem[\protect\citeauthoryear{%
Fern{\'a}ndez%
\ \protect\BOthers{.}}{%
Fern{\'a}ndez%
\ \protect\BOthers{.}}{%
{\protect\APACyear{2009}}%
}]{%
fernandez2009}%
\APACinsertmetastar{%
fernandez2009}%
Fern{\'a}ndez, A.%
, Quintero, J.%
, Hornero, R.%
, Zuluaga, P.%
, Navas, M.%
, G{\'o}mez, C.%
\BCBL{}\ \BOthersPeriod{.}%
\unskip\
\newblock
\APACrefYearMonthDay{2009}{}{}.
\newblock
\BBOQ{}\APACrefatitle{Complexity analysis of spontaneous brain activity in
  attention-deficit/hyperactivity disorder: diagnostic implications}{Complexity
  analysis of spontaneous brain activity in attention-deficit/hyperactivity
  disorder: diagnostic implications}.\BBCQ{}
\newblock
\APACjournalVolNumPages{Biological Psychiatry}{65}{7}{571--577}.
\PrintBackRefs{\CurrentBib}

\bibitem[\protect\citeauthoryear{%
Fern{\'a}ndez%
\ \protect\BOthers{.}}{%
Fern{\'a}ndez%
\ \protect\BOthers{.}}{%
{\protect\APACyear{2011}}%
}]{%
fernandez2011}%
\APACinsertmetastar{%
fernandez2011}%
Fern{\'a}ndez, A.%
, R{\'\i}os-Lago, M.%
, Ab{\'a}solo, D.%
, Hornero, R.%
, {\'A}lvarez-Linera, J.%
, Paul, N.%
\BCBL{}\ \BOthersPeriod{.}%
\unskip\
\newblock
\APACrefYearMonthDay{2011}{}{}.
\newblock
\BBOQ{}\APACrefatitle{The correlation between white-matter microstructure and
  the complexity of spontaneous brain activity: A difussion tensor imaging-MEG
  study}{The correlation between white-matter microstructure and the complexity
  of spontaneous brain activity: A difussion tensor imaging-meg study}.\BBCQ{}
\newblock
\APACjournalVolNumPages{Neuroimage}{57}{4}{1300--1307}.
\PrintBackRefs{\CurrentBib}

\bibitem[\protect\citeauthoryear{%
Fern{\'a}ndez%
\ \protect\BOthers{.}}{%
Fern{\'a}ndez%
\ \protect\BOthers{.}}{%
{\protect\APACyear{2012}}%
}]{%
fernandez2012}%
\APACinsertmetastar{%
fernandez2012}%
Fern{\'a}ndez, A.%
, Zuluaga, P.%
, Ab{\'a}solo, D.%
, G{\'o}mez, C.%
, Serra, A.%
, M{\'e}ndez, M\BPBI A.%
\BCBL{}\ \BOthersPeriod{.}%
\unskip\
\newblock
\APACrefYearMonthDay{2012}{}{}.
\newblock
\BBOQ{}\APACrefatitle{Brain oscillatory complexity across the life span}{Brain
  oscillatory complexity across the life span}.\BBCQ{}
\newblock
\APACjournalVolNumPages{Clinical Neurophysiology}{123}{11}{2154--2162}.
\PrintBackRefs{\CurrentBib}

\bibitem[\protect\citeauthoryear{%
Fournier%
, Amano%
, Radonovich%
, Bleser%
\BCBL{}\ \BBA{} Hass%
}{%
Fournier%
\ \protect\BOthers{.}}{%
{\protect\APACyear{2013}}%
}]{%
fournier2013}%
\APACinsertmetastar{%
fournier2013}%
Fournier, K\BPBI A.%
, Amano, S.%
, Radonovich, K\BPBI J.%
, Bleser, T\BPBI M.%
\BCBL{}\ \BBA{} Hass, C\BPBI J.%
%
\unskip\
\newblock
\APACrefYearMonthDay{2013}{}{}.
\newblock
\BBOQ{}\APACrefatitle{Decreased dynamical complexity during quiet stance in
  children with Autism Spectrum Disorders}{Decreased dynamical complexity
  during quiet stance in children with autism spectrum disorders}.\BBCQ{}
\newblock
\APACjournalVolNumPages{Gait \& Posture}{}{}{}.
\PrintBackRefs{\CurrentBib}

\bibitem[\protect\citeauthoryear{%
{Free Software Foundation}%
}{%
{Free Software Foundation}%
}{%
{\protect\APACyear{2007}}%
}]{%
gplv3}%
\APACinsertmetastar{%
gplv3}%
{Free Software Foundation}.%
%
\unskip\
\newblock
\APACrefYearMonthDay{2007}{}{}.
\newblock
\APACrefbtitle{{GNU} General Public License.}{{GNU} general public license.}
\newblock
 \begin{APACrefURL} \url{http://www.gnu.org/licenses/gpl.html} \end{APACrefURL}
\PrintBackRefs{\CurrentBib}

\bibitem[\protect\citeauthoryear{%
Gauvrit%
, Soler-Toscano%
\BCBL{}\ \BBA{} Zenil%
}{%
Gauvrit%
\ \protect\BOthers{.}}{%
{\protect\APACyear{2014}}%
}]{%
gauvrit2014vision}%
\APACinsertmetastar{%
gauvrit2014vision}%
Gauvrit, N.%
, Soler-Toscano, F.%
\BCBL{}\ \BBA{} Zenil, H.%
%
\unskip\
\newblock
\APACrefYearMonthDay{2014}{}{}.
\newblock
\BBOQ{}\APACrefatitle{Natural scene statistics mediate the perception of image
  complexity}{Natural scene statistics mediate the perception of image
  complexity}.\BBCQ{}
\newblock
\APACjournalVolNumPages{Visual Cognition}{22}{8}{1084-1091}.
\PrintBackRefs{\CurrentBib}

\bibitem[\protect\citeauthoryear{%
Gauvrit%
, Zenil%
, Delahaye%
\BCBL{}\ \BBA{} Soler-Toscano%
}{%
Gauvrit%
\ \protect\BOthers{.}}{%
{\protect\APACyear{2013}}%
}]{%
gauvrit2014}%
\APACinsertmetastar{%
gauvrit2014}%
Gauvrit, N.%
, Zenil, H.%
, Delahaye, J\BHBI P.%
\BCBL{}\ \BBA{} Soler-Toscano, F.%
%
\unskip\
\newblock
\APACrefYearMonthDay{2013}{}{}.
\newblock
\BBOQ{}\APACrefatitle{Algorithmic complexity for short binary strings applied
  to psychology: {A} primer}{Algorithmic complexity for short binary strings
  applied to psychology: {A} primer}.\BBCQ{}
\newblock
\APACjournalVolNumPages{Behavior Research Methods}{46}{3}{732-744}.
\PrintBackRefs{\CurrentBib}

\bibitem[\protect\citeauthoryear{%
Griffiths%
\ \BBA{} Tenenbaum%
}{%
Griffiths%
\ \BBA{} Tenenbaum%
}{%
{\protect\APACyear{2003}}%
}]{%
griffiths2003}%
\APACinsertmetastar{%
griffiths2003}%
Griffiths, T\BPBI L.%
\BCBT{}\ \BBA{} Tenenbaum, J\BPBI B.%
%
\unskip\
\newblock
\APACrefYearMonthDay{2003}{}{}.
\newblock
\BBOQ{}\APACrefatitle{Probability, algorithmic complexity, and subjective
  randomness}{Probability, algorithmic complexity, and subjective
  randomness}.\BBCQ{}
\newblock
\BIn{} R.~Alterman\ \BBA{} D.~Kirsch\ (\BEDS), \APACrefbtitle{Proceedings of
  the 25th Annual Conference of the Cognitive Science Society}{Proceedings of
  the 25th annual conference of the cognitive science society}\ (\BPG~480-485).
\newblock
\APACaddressPublisher{Mahwah, NJ}{Erlbaum}.
\PrintBackRefs{\CurrentBib}

\bibitem[\protect\citeauthoryear{%
Griffiths%
\ \BBA{} Tenenbaum%
}{%
Griffiths%
\ \BBA{} Tenenbaum%
}{%
{\protect\APACyear{2004}}%
}]{%
griffiths2004}%
\APACinsertmetastar{%
griffiths2004}%
Griffiths, T\BPBI L.%
\BCBT{}\ \BBA{} Tenenbaum, J\BPBI B.%
%
\unskip\
\newblock
\APACrefYearMonthDay{2004}{}{}.
\newblock
\BBOQ{}\APACrefatitle{From algorithmic to sub- jective randomness}{From
  algorithmic to sub- jective randomness}.\BBCQ{}
\newblock
\BIn{} S.~Thrun, L\BPBI K.~Saul\BCBL{}\ \BBA{} B.~Sch{\"o}lkopf\ (\BEDS),
  \APACrefbtitle{Advances in neural information processing systems}{Advances in
  neural information processing systems}\ (\BVOL~16, \BPG~953-960).
\newblock
\APACaddressPublisher{Cambridge, MA}{MIT Press}.
\PrintBackRefs{\CurrentBib}

\bibitem[\protect\citeauthoryear{%
Gruber%
}{%
Gruber%
}{%
{\protect\APACyear{2010}}%
}]{%
gruber2010}%
\APACinsertmetastar{%
gruber2010}%
Gruber, H.%
%
\unskip\
\newblock
\APACrefYear{2010}.
\newblock
\APACrefbtitle{On the descriptional and algorithmic complexity of regular
  languages}{On the descriptional and algorithmic complexity of regular
  languages}.
\newblock
\APACaddressPublisher{}{Justus Liebig University Giessen}.
\PrintBackRefs{\CurrentBib}

\bibitem[\protect\citeauthoryear{%
Gr{\"u}nwald%
}{%
Gr{\"u}nwald%
}{%
{\protect\APACyear{2007}}%
}]{%
grunwald2007minimum}%
\APACinsertmetastar{%
grunwald2007minimum}%
Gr{\"u}nwald, P\BPBI D.%
%
\unskip\
\newblock
\APACrefYear{2007}.
\newblock
\APACrefbtitle{The minimum description length principle}{The minimum
  description length principle}.
\newblock
\APACaddressPublisher{}{MIT press}.
\PrintBackRefs{\CurrentBib}

\bibitem[\protect\citeauthoryear{%
T.~Hahn%
\ \protect\BOthers{.}}{%
T.~Hahn%
\ \protect\BOthers{.}}{%
{\protect\APACyear{2012}}%
}]{%
hahn2012}%
\APACinsertmetastar{%
hahn2012}%
Hahn, T.%
, Dresler, T.%
, Ehlis, A\BHBI C.%
, Pyka, M.%
, Dieler, A\BPBI C.%
, Saathoff, C.%
\BCBL{}\ \BOthersPeriod{.}%
\unskip\
\newblock
\APACrefYearMonthDay{2012}{}{}.
\newblock
\BBOQ{}\APACrefatitle{Randomness of resting-state brain oscillations encodes
  Gray's personality trait}{Randomness of resting-state brain oscillations
  encodes gray's personality trait}.\BBCQ{}
\newblock
\APACjournalVolNumPages{Neuroimage}{59}{2}{1842--1845}.
\PrintBackRefs{\CurrentBib}

\bibitem[\protect\citeauthoryear{%
U.~Hahn%
}{%
U.~Hahn%
}{%
{\protect\APACyear{2014}}%
}]{%
hahn_experiential_2014}%
\APACinsertmetastar{%
hahn_experiential_2014}%
Hahn, U.%
%
\unskip\
\newblock
\APACrefYearMonthDay{2014}{}{}.
\newblock
\BBOQ{}\APACrefatitle{Experiential Limitation in Judgment and
  Decision}{Experiential limitation in judgment and decision}.\BBCQ{}
\newblock
\APACjournalVolNumPages{Topics in Cognitive Science}{6}{2}{229--244}.
\PrintBackRefs{\CurrentBib}

\bibitem[\protect\citeauthoryear{%
U.~Hahn%
, Chater%
\BCBL{}\ \BBA{} Richardson%
}{%
U.~Hahn%
\ \protect\BOthers{.}}{%
{\protect\APACyear{2003}}%
}]{%
hahn2003}%
\APACinsertmetastar{%
hahn2003}%
Hahn, U.%
, Chater, N.%
\BCBL{}\ \BBA{} Richardson, L\BPBI B.%
%
\unskip\
\newblock
\APACrefYearMonthDay{2003}{}{}.
\newblock
\BBOQ{}\APACrefatitle{Similarity as transformation}{Similarity as
  transformation}.\BBCQ{}
\newblock
\APACjournalVolNumPages{Cognition}{87}{1}{1-32}.
\PrintBackRefs{\CurrentBib}

\bibitem[\protect\citeauthoryear{%
U.~Hahn%
\ \BBA{} Warren%
}{%
U.~Hahn%
\ \BBA{} Warren%
}{%
{\protect\APACyear{2009}}%
}]{%
hahn_perceptions_2009}%
\APACinsertmetastar{%
hahn_perceptions_2009}%
Hahn, U.%
\BCBT{}\ \BBA{} Warren, P\BPBI A.%
%
\unskip\
\newblock
\APACrefYearMonthDay{2009}{}{}.
\newblock
\BBOQ{}\APACrefatitle{Perceptions of randomness: Why three heads are better
  than four.}{Perceptions of randomness: Why three heads are better than
  four.}\BBCQ{}
\newblock
\APACjournalVolNumPages{Psychological Review}{116}{2}{454--461}.
\PrintBackRefs{\CurrentBib}

\bibitem[\protect\citeauthoryear{%
Heuer%
, Kohlisch%
\BCBL{}\ \BBA{} Klein%
}{%
Heuer%
\ \protect\BOthers{.}}{%
{\protect\APACyear{2005}}%
}]{%
heuer2005sleep}%
\APACinsertmetastar{%
heuer2005sleep}%
Heuer, H.%
, Kohlisch, O.%
\BCBL{}\ \BBA{} Klein, W.%
%
\unskip\
\newblock
\APACrefYearMonthDay{2005}{}{}.
\newblock
\BBOQ{}\APACrefatitle{The effects of total sleep deprivation on the generation
  of random sequences of key-presses, numbers and nouns.}{The effects of total
  sleep deprivation on the generation of random sequences of key-presses,
  numbers and nouns.}\BBCQ{}
\newblock
\APACjournalVolNumPages{The Quarterly Journal of Experimental Psychology A:
  Human Experimental Psychology}{58A}{2}{275 - 307}.
\PrintBackRefs{\CurrentBib}

\bibitem[\protect\citeauthoryear{%
Hsu%
, Griffiths%
\BCBL{}\ \BBA{} Schreiber%
}{%
Hsu%
\ \protect\BOthers{.}}{%
{\protect\APACyear{2010}}%
}]{%
hsu2010subjective}%
\APACinsertmetastar{%
hsu2010subjective}%
Hsu, A\BPBI S.%
, Griffiths, T\BPBI L.%
\BCBL{}\ \BBA{} Schreiber, E.%
%
\unskip\
\newblock
\APACrefYearMonthDay{2010}{}{}.
\newblock
\BBOQ{}\APACrefatitle{Subjective randomness and natural scene
  statistics}{Subjective randomness and natural scene statistics}.\BBCQ{}
\newblock
\APACjournalVolNumPages{Psychonomic Bulletin \& Review}{17}{5}{624--629}.
\PrintBackRefs{\CurrentBib}

\bibitem[\protect\citeauthoryear{%
Jones%
, Maillardet%
\BCBL{}\ \BBA{} Robinson%
}{%
Jones%
\ \protect\BOthers{.}}{%
{\protect\APACyear{2009}}%
}]{%
jones_introduction_2009}%
\APACinsertmetastar{%
jones_introduction_2009}%
Jones, O.%
, Maillardet, R.%
\BCBL{}\ \BBA{} Robinson, A.%
%
\unskip\
\newblock
\APACrefYear{2009}.
\newblock
\APACrefbtitle{Introduction to scientific programming and simulation using
  {R}}{Introduction to scientific programming and simulation using {R}}.
\newblock
\APACaddressPublisher{Boca Raton, {FL}}{Chapman \& {Hall/CRC}}.
\PrintBackRefs{\CurrentBib}

\bibitem[\protect\citeauthoryear{%
Kahneman%
, Slovic%
\BCBL{}\ \BBA{} Tversky%
}{%
Kahneman%
\ \protect\BOthers{.}}{%
{\protect\APACyear{1982}}%
}]{%
kahneman1982judgment}%
\APACinsertmetastar{%
kahneman1982judgment}%
Kahneman, D.%
, Slovic, P.%
\BCBL{}\ \BBA{} Tversky, A.%
%
\unskip\
\newblock
\APACrefYear{1982}.
\newblock
\APACrefbtitle{Judgment under uncertainty: Heuristics and biases}{Judgment
  under uncertainty: Heuristics and biases}.
\newblock
\APACaddressPublisher{}{Cambridge University Press}.
\PrintBackRefs{\CurrentBib}

\bibitem[\protect\citeauthoryear{%
Kass%
\ \BBA{} Raftery%
}{%
Kass%
\ \BBA{} Raftery%
}{%
{\protect\APACyear{1995}}%
}]{%
kass_bayes_1995}%
\APACinsertmetastar{%
kass_bayes_1995}%
Kass, R\BPBI E.%
\BCBT{}\ \BBA{} Raftery, A\BPBI E.%
%
\unskip\
\newblock
\APACrefYearMonthDay{1995}{{\APACmonth{06}}}{}.
\newblock
\BBOQ{}\APACrefatitle{Bayes Factors}{Bayes factors}.\BBCQ{}
\newblock
\APACjournalVolNumPages{Journal of the American Statistical
  Association}{90}{430}{773--795}.
\newblock
 \begin{APACrefURL}
  \url{http://www.tandfonline.com/doi/abs/10.1080/01621459.1995.10476572}
  \end{APACrefURL}
\PrintBackRefs{\CurrentBib}

\bibitem[\protect\citeauthoryear{%
Kellen%
, Klauer%
\BCBL{}\ \BBA{} Br\"oder%
}{%
Kellen%
\ \protect\BOthers{.}}{%
{\protect\APACyear{2013}}%
}]{%
kellen_recognition_2013}%
\APACinsertmetastar{%
kellen_recognition_2013}%
Kellen, D.%
, Klauer, K\BPBI C.%
\BCBL{}\ \BBA{} Br\"oder, A.%
%
\unskip\
\newblock
\APACrefYearMonthDay{2013}{}{}.
\newblock
\BBOQ{}\APACrefatitle{Recognition memory models and binary-response {ROCs:} A
  comparison by minimum description length}{Recognition memory models and
  binary-response {ROCs:} a comparison by minimum description length}.\BBCQ{}
\newblock
\APACjournalVolNumPages{Psychonomic Bulletin \& Review}{20}{4}{693--719}.
\PrintBackRefs{\CurrentBib}

\bibitem[\protect\citeauthoryear{%
Koike%
\ \protect\BOthers{.}}{%
Koike%
\ \protect\BOthers{.}}{%
{\protect\APACyear{2011}}%
}]{%
koike2011}%
\APACinsertmetastar{%
koike2011}%
Koike, S.%
, Takizawa, R.%
, Nishimura, Y.%
, Marumo, K.%
, Kinou, M.%
, Kawakubo, Y.%
\BCBL{}\ \BOthersPeriod{.}%
\unskip\
\newblock
\APACrefYearMonthDay{2011}{}{}.
\newblock
\BBOQ{}\APACrefatitle{Association between severe dorsolateral prefrontal
  dysfunction during random number generation and earlier onset in
  schizophrenia}{Association between severe dorsolateral prefrontal dysfunction
  during random number generation and earlier onset in schizophrenia}.\BBCQ{}
\newblock
\APACjournalVolNumPages{Clinical Neurophysiology}{122}{8}{1533--1540}.
\PrintBackRefs{\CurrentBib}

\bibitem[\protect\citeauthoryear{%
Kolmogorov%
}{%
Kolmogorov%
}{%
{\protect\APACyear{1965}}%
}]{%
kolmogorov1965}%
\APACinsertmetastar{%
kolmogorov1965}%
Kolmogorov, A.%
%
\unskip\
\newblock
\APACrefYearMonthDay{1965}{}{}.
\newblock
\BBOQ{}\APACrefatitle{Three approaches to the quantitative definition of
  information}{Three approaches to the quantitative definition of
  information}.\BBCQ{}
\newblock
\APACjournalVolNumPages{Problems of Information and Transmission}{1}{1}{1-7}.
\PrintBackRefs{\CurrentBib}

\bibitem[\protect\citeauthoryear{%
Lai%
\ \protect\BOthers{.}}{%
Lai%
\ \protect\BOthers{.}}{%
{\protect\APACyear{2010}}%
}]{%
lai2010}%
\APACinsertmetastar{%
lai2010}%
Lai, M\BHBI C.%
, Lombardo, M\BPBI V.%
, Chakrabarti, B.%
, Sadek, S\BPBI A.%
, Pasco, G.%
, Wheelwright, S\BPBI J.%
\BCBL{}\ \BOthersPeriod{.}%
\unskip\
\newblock
\APACrefYearMonthDay{2010}{}{}.
\newblock
\BBOQ{}\APACrefatitle{A shift to randomness of brain oscillations in people
  with autism}{A shift to randomness of brain oscillations in people with
  autism}.\BBCQ{}
\newblock
\APACjournalVolNumPages{Biological Psychiatry}{68}{12}{1092--1099}.
\PrintBackRefs{\CurrentBib}

\bibitem[\protect\citeauthoryear{%
Levin%
}{%
Levin%
}{%
{\protect\APACyear{1974}}%
}]{%
levin1974}%
\APACinsertmetastar{%
levin1974}%
Levin, L\BPBI A.%
%
\unskip\
\newblock
\APACrefYearMonthDay{1974}{}{}.
\newblock
\BBOQ{}\APACrefatitle{Laws of information conservation (nongrowth) and aspects
  of the foundation of probability theory}{Laws of information conservation
  (nongrowth) and aspects of the foundation of probability theory}.\BBCQ{}
\newblock
\APACjournalVolNumPages{Problemy Peredachi Informatsii}{10}{3}{30--35}.
\PrintBackRefs{\CurrentBib}

\bibitem[\protect\citeauthoryear{%
Li%
\ \BBA{} Vit\'anyi%
}{%
Li%
\ \BBA{} Vit\'anyi%
}{%
{\protect\APACyear{2008}}%
}]{%
li2008}%
\APACinsertmetastar{%
li2008}%
Li, M.%
\BCBT{}\ \BBA{} Vit\'anyi, P.%
%
\unskip\
\newblock
\APACrefYear{2008}.
\newblock
\APACrefbtitle{An Introduction to Kolmogorov Complexity and Its
  Applications}{An introduction to kolmogorov complexity and its applications}.
\newblock
\APACaddressPublisher{}{Springer Verlag}.
\PrintBackRefs{\CurrentBib}

\bibitem[\protect\citeauthoryear{%
Loetscher%
\ \BBA{} Brugger%
}{%
Loetscher%
\ \BBA{} Brugger%
}{%
{\protect\APACyear{2009}}%
}]{%
loetscher2009neglect}%
\APACinsertmetastar{%
loetscher2009neglect}%
Loetscher, T.%
\BCBT{}\ \BBA{} Brugger, P.%
%
\unskip\
\newblock
\APACrefYearMonthDay{2009}{}{}.
\newblock
\BBOQ{}\APACrefatitle{Random number generation in neglect patients reveals
  enhanced response stereotypy, but no neglect in number space.}{Random number
  generation in neglect patients reveals enhanced response stereotypy, but no
  neglect in number space.}\BBCQ{}
\newblock
\APACjournalVolNumPages{Neuropsychologia}{47}{1}{276 - 279}.
\PrintBackRefs{\CurrentBib}

\bibitem[\protect\citeauthoryear{%
Machado%
, Miranda%
, Morya%
, Amaro~Jr%
\BCBL{}\ \BBA{} Sameshima%
}{%
Machado%
\ \protect\BOthers{.}}{%
{\protect\APACyear{2010}}%
}]{%
machado2010}%
\APACinsertmetastar{%
machado2010}%
Machado, B.%
, Miranda, T.%
, Morya, E.%
, Amaro~Jr, E.%
\BCBL{}\ \BBA{} Sameshima, K.%
%
\unskip\
\newblock
\APACrefYearMonthDay{2010}{}{}.
\newblock
\BBOQ{}\APACrefatitle{P24-23 Algorithmic complexity measure of {EEG} for
  staging brain state}{P24-23 algorithmic complexity measure of {EEG} for
  staging brain state}.\BBCQ{}
\newblock
\APACjournalVolNumPages{Clinical Neurophysiology}{121}{}{S249--S250}.
\PrintBackRefs{\CurrentBib}

\bibitem[\protect\citeauthoryear{%
Maes%
, Vissers%
, Egger%
\BCBL{}\ \BBA{} Eling%
}{%
Maes%
\ \protect\BOthers{.}}{%
{\protect\APACyear{2012}}%
}]{%
maes2012}%
\APACinsertmetastar{%
maes2012}%
Maes, J\BPBI H.%
, Vissers, C\BPBI T.%
, Egger, J\BPBI I.%
\BCBL{}\ \BBA{} Eling, P\BPBI A.%
%
\unskip\
\newblock
\APACrefYearMonthDay{2012}{}{}.
\newblock
\BBOQ{}\APACrefatitle{On the relationship between autistic traits and executive
  functioning in a non-clinical {D}utch student population}{On the relationship
  between autistic traits and executive functioning in a non-clinical {D}utch
  student population}.\BBCQ{}
\newblock
\APACjournalVolNumPages{Autism}{17}{4}{379--389}.
\PrintBackRefs{\CurrentBib}

\bibitem[\protect\citeauthoryear{%
Maindonald%
\ \BBA{} Braun%
}{%
Maindonald%
\ \BBA{} Braun%
}{%
{\protect\APACyear{2010}}%
}]{%
maindonald_data_2010}%
\APACinsertmetastar{%
maindonald_data_2010}%
Maindonald, J.%
\BCBT{}\ \BBA{} Braun, W\BPBI J.%
%
\unskip\
\newblock
\APACrefYear{2010}.
\newblock
\APACrefbtitle{Data Analysis and Graphics Using {R}: An Example-Based
  Approach}{Data analysis and graphics using {R}: An example-based approach}\
  (\PrintOrdinal{3}\ \BEd).
\newblock
\APACaddressPublisher{Cambridge}{Cambridge University Press}.
\PrintBackRefs{\CurrentBib}

\bibitem[\protect\citeauthoryear{%
Manktelow%
\ \BBA{} Over%
}{%
Manktelow%
\ \BBA{} Over%
}{%
{\protect\APACyear{1993}}%
}]{%
manktelow1993rationality}%
\APACinsertmetastar{%
manktelow1993rationality}%
Manktelow, K\BPBI I.%
\BCBT{}\ \BBA{} Over, D\BPBI E.%
%
\unskip\
\newblock
\APACrefYear{1993}.
\newblock
\APACrefbtitle{Rationality: Psychological and philosophical
  perspectives.}{Rationality: Psychological and philosophical perspectives.}
\newblock
\APACaddressPublisher{}{Taylor \& Frances/Routledge}.
\PrintBackRefs{\CurrentBib}

\bibitem[\protect\citeauthoryear{%
Martin-L{\"o}f%
}{%
Martin-L{\"o}f%
}{%
{\protect\APACyear{1966}}%
}]{%
martin1966}%
\APACinsertmetastar{%
martin1966}%
Martin-L{\"o}f, P.%
%
\unskip\
\newblock
\APACrefYearMonthDay{1966}{}{}.
\newblock
\BBOQ{}\APACrefatitle{The definition of random sequences}{The definition of
  random sequences}.\BBCQ{}
\newblock
\APACjournalVolNumPages{Information and control}{9}{6}{602--619}.
\PrintBackRefs{\CurrentBib}

\bibitem[\protect\citeauthoryear{%
Mathy%
\ \BBA{} Feldman%
}{%
Mathy%
\ \BBA{} Feldman%
}{%
{\protect\APACyear{2012}}%
}]{%
mathy2012}%
\APACinsertmetastar{%
mathy2012}%
Mathy, F.%
\BCBT{}\ \BBA{} Feldman, J.%
%
\unskip\
\newblock
\APACrefYearMonthDay{2012}{}{}.
\newblock
\BBOQ{}\APACrefatitle{What's magic about magic numbers? {C}hunking and data
  compression in short-term memory}{What's magic about magic numbers?
  {C}hunking and data compression in short-term memory}.\BBCQ{}
\newblock
\APACjournalVolNumPages{Cognition}{122}{3}{346--362}.
\PrintBackRefs{\CurrentBib}

\bibitem[\protect\citeauthoryear{%
Matloff%
}{%
Matloff%
}{%
{\protect\APACyear{2011}}%
}]{%
matloff_art_2011}%
\APACinsertmetastar{%
matloff_art_2011}%
Matloff, N.%
%
\unskip\
\newblock
\APACrefYear{2011}.
\newblock
\APACrefbtitle{The Art of {R} Programming: A Tour of Statistical Software
  Design}{The art of {R} programming: A tour of statistical software design}\
  (\PrintOrdinal{1}\ \BEd).
\newblock
\APACaddressPublisher{San Francisco}{No Starch Press}.
\PrintBackRefs{\CurrentBib}

\bibitem[\protect\citeauthoryear{%
Matthews%
}{%
Matthews%
}{%
{\protect\APACyear{2013}}%
}]{%
matthews2013}%
\APACinsertmetastar{%
matthews2013}%
Matthews, W.%
%
\unskip\
\newblock
\APACrefYearMonthDay{2013}{}{}.
\newblock
\BBOQ{}\APACrefatitle{Relatively Random: Context Effects on Perceived
  Randomness and Predicted Outcomes}{Relatively random: Context effects on
  perceived randomness and predicted outcomes}.\BBCQ{}
\newblock
\APACjournalVolNumPages{Journal of Experimental Psychology: Learning, Memory,
  and Cognition}{39}{5}{1642-1648}.
\PrintBackRefs{\CurrentBib}

\bibitem[\protect\citeauthoryear{%
Miller%
}{%
Miller%
}{%
{\protect\APACyear{1956}}%
}]{%
miller1956magical}%
\APACinsertmetastar{%
miller1956magical}%
Miller, G\BPBI A.%
%
\unskip\
\newblock
\APACrefYearMonthDay{1956}{}{}.
\newblock
\BBOQ{}\APACrefatitle{The magical number seven, plus or minus two: some limits
  on our capacity for processing information.}{The magical number seven, plus
  or minus two: some limits on our capacity for processing information.}\BBCQ{}
\newblock
\APACjournalVolNumPages{Psychological Review}{63}{2}{81--97}.
\PrintBackRefs{\CurrentBib}

\bibitem[\protect\citeauthoryear{%
Myung%
, Cavagnaro%
\BCBL{}\ \BBA{} Pitt%
}{%
Myung%
\ \protect\BOthers{.}}{%
{\protect\APACyear{In press}}%
}]{%
myung2014}%
\APACinsertmetastar{%
myung2014}%
Myung, J\BPBI I.%
, Cavagnaro, D\BPBI R.%
\BCBL{}\ \BBA{} Pitt, M\BPBI A.%
%
\unskip\
\newblock
\APACrefYearMonthDay{In press}{}{}.
\newblock
\BBOQ{}\APACrefatitle{New Handbook of Mathematical Psychology, Vol. 1:
  Measurement and Methodology}{New handbook of mathematical psychology, vol. 1:
  Measurement and methodology}.\BBCQ{}
\newblock
\BIn{} W\BPBI H.~Batchelder, H.~Colonius, E.~Dzhafarov\BCBL{}\ \BBA{} J\BPBI
  I.~Myung\ (\BEDS), (\BCHAP\ Model evaluation and selection).
\newblock
\APACaddressPublisher{}{Cambridge University Press}.
\PrintBackRefs{\CurrentBib}

\bibitem[\protect\citeauthoryear{%
Myung%
, Navarro%
\BCBL{}\ \BBA{} Pitt%
}{%
Myung%
\ \protect\BOthers{.}}{%
{\protect\APACyear{2006}}%
}]{%
myung_model_2006}%
\APACinsertmetastar{%
myung_model_2006}%
Myung, J\BPBI I.%
, Navarro, D\BPBI J.%
\BCBL{}\ \BBA{} Pitt, M\BPBI A.%
%
\unskip\
\newblock
\APACrefYearMonthDay{2006}{}{}.
\newblock
\BBOQ{}\APACrefatitle{Model selection by normalized maximum likelihood}{Model
  selection by normalized maximum likelihood}.\BBCQ{}
\newblock
\APACjournalVolNumPages{Journal of Mathematical Psychology}{50}{2}{167--179}.
\PrintBackRefs{\CurrentBib}

\bibitem[\protect\citeauthoryear{%
Naranan%
}{%
Naranan%
}{%
{\protect\APACyear{2011}}%
}]{%
naranan2011}%
\APACinsertmetastar{%
naranan2011}%
Naranan, S.%
%
\unskip\
\newblock
\APACrefYearMonthDay{2011}{}{}.
\newblock
\BBOQ{}\APACrefatitle{Historical Linguistics and Evolutionary Genetics. Based
  on Symbol Frequencies in Tamil Texts and DNA Sequences}{Historical
  linguistics and evolutionary genetics. based on symbol frequencies in tamil
  texts and dna sequences}.\BBCQ{}
\newblock
\APACjournalVolNumPages{Journal of Quantitative Linguistics}{18}{4}{337--358}.
\PrintBackRefs{\CurrentBib}

\bibitem[\protect\citeauthoryear{%
Nies%
}{%
Nies%
}{%
{\protect\APACyear{2009}}%
}]{%
nies2009}%
\APACinsertmetastar{%
nies2009}%
Nies, A.%
%
\unskip\
\newblock
\APACrefYear{2009}.
\newblock
\APACrefbtitle{Computability and randomness}{Computability and randomness}\
  (\BVOL~51).
\newblock
\APACaddressPublisher{}{Oxford University Press}.
\PrintBackRefs{\CurrentBib}

\bibitem[\protect\citeauthoryear{%
Over%
}{%
Over%
}{%
{\protect\APACyear{2009}}%
}]{%
over09}%
\APACinsertmetastar{%
over09}%
Over, D\BPBI E.%
%
\unskip\
\newblock
\APACrefYearMonthDay{2009}{}{}.
\newblock
\BBOQ{}\APACrefatitle{New paradigm psychology of reasoning}{New paradigm
  psychology of reasoning}.\BBCQ{}
\newblock
\APACjournalVolNumPages{Thinking \& Reasoning}{15}{4}{431--438}.
\PrintBackRefs{\CurrentBib}

\bibitem[\protect\citeauthoryear{%
Pearson%
\ \BBA{} Sawyer%
}{%
Pearson%
\ \BBA{} Sawyer%
}{%
{\protect\APACyear{2011}}%
}]{%
pearson2011}%
\APACinsertmetastar{%
pearson2011}%
Pearson, D\BPBI G.%
\BCBT{}\ \BBA{} Sawyer, T.%
%
\unskip\
\newblock
\APACrefYearMonthDay{2011}{}{}.
\newblock
\BBOQ{}\APACrefatitle{Effects of dual task interference on memory intrusions
  for affective images}{Effects of dual task interference on memory intrusions
  for affective images}.\BBCQ{}
\newblock
\APACjournalVolNumPages{International Journal of Cognitive
  Therapy}{4}{2}{122--133}.
\PrintBackRefs{\CurrentBib}

\bibitem[\protect\citeauthoryear{%
Proios%
, Asaridou%
\BCBL{}\ \BBA{} Brugger%
}{%
Proios%
\ \protect\BOthers{.}}{%
{\protect\APACyear{2008}}%
}]{%
Proios08}%
\APACinsertmetastar{%
Proios08}%
Proios, H.%
, Asaridou, S\BPBI S.%
\BCBL{}\ \BBA{} Brugger, P.%
%
\unskip\
\newblock
\APACrefYearMonthDay{2008}{}{}.
\newblock
\BBOQ{}\APACrefatitle{Random number generation in patients with aphasia: A test
  of executive functions}{Random number generation in patients with aphasia: A
  test of executive functions}.\BBCQ{}
\newblock
\APACjournalVolNumPages{Acta Neuropsychologica}{6}{}{157-168}.
\PrintBackRefs{\CurrentBib}

\bibitem[\protect\citeauthoryear{%
Pureza%
, Gon{\c{c}}alves%
, Branco%
, Grassi-Oliveira%
\BCBL{}\ \BBA{} Fonseca%
}{%
Pureza%
\ \protect\BOthers{.}}{%
{\protect\APACyear{2013}}%
}]{%
pureza2013}%
\APACinsertmetastar{%
pureza2013}%
Pureza, J\BPBI R.%
, Gon{\c{c}}alves, H\BPBI A.%
, Branco, L.%
, Grassi-Oliveira, R.%
\BCBL{}\ \BBA{} Fonseca, R\BPBI P.%
%
\unskip\
\newblock
\APACrefYearMonthDay{2013}{}{}.
\newblock
\BBOQ{}\APACrefatitle{Executive functions in late childhood: Age differences
  among groups}{Executive functions in late childhood: Age differences among
  groups}.\BBCQ{}
\newblock
\APACjournalVolNumPages{Psychology \& Neuroscience}{6}{1}{79--88}.
\PrintBackRefs{\CurrentBib}

\bibitem[\protect\citeauthoryear{%
{R Core Team}%
}{%
{R Core Team}%
}{%
{\protect\APACyear{2014}}%
}]{%
r2014}%
\APACinsertmetastar{%
r2014}%
{R Core Team}.%
%
\unskip\
\newblock
\APACrefYearMonthDay{2014}{}{}.
\newblock
\BBOQ{}\APACrefatitle{R: A Language and Environment for Statistical
  Computing}{R: A language and environment for statistical computing}\BBCQ{}\
  [\bibcomputersoftwaremanual].
\newblock
\APACaddressPublisher{Vienna, Austria}{}.
\newblock
 \begin{APACrefURL} \url{http://www.R-project.org/} \end{APACrefURL}
\PrintBackRefs{\CurrentBib}

\bibitem[\protect\citeauthoryear{%
Rado%
}{%
Rado%
}{%
{\protect\APACyear{1962}}%
}]{%
rado1962}%
\APACinsertmetastar{%
rado1962}%
Rado, T.%
%
\unskip\
\newblock
\APACrefYearMonthDay{1962}{}{}.
\newblock
\BBOQ{}\APACrefatitle{On Non-Computable Functions}{On non-computable
  functions}.\BBCQ{}
\newblock
\APACjournalVolNumPages{Bell System Technical Journal}{41}{}{877-884}.
\PrintBackRefs{\CurrentBib}

\bibitem[\protect\citeauthoryear{%
Rissanen%
}{%
Rissanen%
}{%
{\protect\APACyear{1989}}%
}]{%
rissanen1989stochastic}%
\APACinsertmetastar{%
rissanen1989stochastic}%
Rissanen, J.%
%
\unskip\
\newblock
\APACrefYear{1989}.
\newblock
\APACrefbtitle{Stochastic complexity in statistical inquiry theory}{Stochastic
  complexity in statistical inquiry theory}.
\newblock
\APACaddressPublisher{}{World Scientific Publishing Co., Inc.}
\PrintBackRefs{\CurrentBib}

\bibitem[\protect\citeauthoryear{%
Ryabko%
, Reznikova%
, Druzyaka%
\BCBL{}\ \BBA{} Panteleeva%
}{%
Ryabko%
\ \protect\BOthers{.}}{%
{\protect\APACyear{2013}}%
}]{%
ryabko2013}%
\APACinsertmetastar{%
ryabko2013}%
Ryabko, B.%
, Reznikova, Z.%
, Druzyaka, A.%
\BCBL{}\ \BBA{} Panteleeva, S.%
%
\unskip\
\newblock
\APACrefYearMonthDay{2013}{}{}.
\newblock
\BBOQ{}\APACrefatitle{Using Ideas of {K}olmogorov Complexity for Studying
  Biological Texts}{Using ideas of {K}olmogorov complexity for studying
  biological texts}.\BBCQ{}
\newblock
\APACjournalVolNumPages{Theory of Computing Systems}{52}{1}{133--147}.
\PrintBackRefs{\CurrentBib}

\bibitem[\protect\citeauthoryear{%
Scafetta%
, Marchi%
\BCBL{}\ \BBA{} West%
}{%
Scafetta%
\ \protect\BOthers{.}}{%
{\protect\APACyear{2009}}%
}]{%
scafetta2009}%
\APACinsertmetastar{%
scafetta2009}%
Scafetta, N.%
, Marchi, D.%
\BCBL{}\ \BBA{} West, B\BPBI J.%
%
\unskip\
\newblock
\APACrefYearMonthDay{2009}{}{}.
\newblock
\BBOQ{}\APACrefatitle{Understanding the complexity of human gait
  dynamics}{Understanding the complexity of human gait dynamics}.\BBCQ{}
\newblock
\APACjournalVolNumPages{Chaos: An Interdisciplinary Journal of Nonlinear
  Science}{19}{2}{026108}.
\PrintBackRefs{\CurrentBib}

\bibitem[\protect\citeauthoryear{%
Schnorr%
}{%
Schnorr%
}{%
{\protect\APACyear{1973}}%
}]{%
schnorr1973}%
\APACinsertmetastar{%
schnorr1973}%
Schnorr, C\BHBI P.%
%
\unskip\
\newblock
\APACrefYearMonthDay{1973}{}{}.
\newblock
\BBOQ{}\APACrefatitle{Process complexity and effective random tests}{Process
  complexity and effective random tests}.\BBCQ{}
\newblock
\APACjournalVolNumPages{Journal of Computer and System
  Sciences}{7}{4}{376--388}.
\PrintBackRefs{\CurrentBib}

\bibitem[\protect\citeauthoryear{%
Schulter%
, Mittenecker%
\BCBL{}\ \BBA{} Papousek%
}{%
Schulter%
\ \protect\BOthers{.}}{%
{\protect\APACyear{2010}}%
}]{%
schulter2010mittenecker}%
\APACinsertmetastar{%
schulter2010mittenecker}%
Schulter, G.%
, Mittenecker, E.%
\BCBL{}\ \BBA{} Papousek, I.%
%
\unskip\
\newblock
\APACrefYearMonthDay{2010}{}{}.
\newblock
\BBOQ{}\APACrefatitle{A computer program for testing and analyzing random
  generation behavior in normal and clinical samples: The {M}ittenecker
  Pointing Test}{A computer program for testing and analyzing random generation
  behavior in normal and clinical samples: The {M}ittenecker pointing
  test}.\BBCQ{}
\newblock
\APACjournalVolNumPages{Behavior Research Methods}{42}{}{333-341}.
\PrintBackRefs{\CurrentBib}

\bibitem[\protect\citeauthoryear{%
Scibinetti%
, Tocci%
\BCBL{}\ \BBA{} Pesce%
}{%
Scibinetti%
\ \protect\BOthers{.}}{%
{\protect\APACyear{2011}}%
}]{%
scibinetti2011}%
\APACinsertmetastar{%
scibinetti2011}%
Scibinetti, P.%
, Tocci, N.%
\BCBL{}\ \BBA{} Pesce, C.%
%
\unskip\
\newblock
\APACrefYearMonthDay{2011}{}{}.
\newblock
\BBOQ{}\APACrefatitle{Motor Creativity and Creative Thinking in Children: The
  Diverging Role of Inhibition}{Motor creativity and creative thinking in
  children: The diverging role of inhibition}.\BBCQ{}
\newblock
\APACjournalVolNumPages{Creativity Research Journal}{23}{3}{262--272}.
\PrintBackRefs{\CurrentBib}

\bibitem[\protect\citeauthoryear{%
Shannon%
}{%
Shannon%
}{%
{\protect\APACyear{1948}}%
}]{%
shannon1948}%
\APACinsertmetastar{%
shannon1948}%
Shannon, C\BPBI E.%
%
\unskip\
\newblock
\APACrefYearMonthDay{1948}{}{}.
\newblock
\BBOQ{}\APACrefatitle{A mathematical theory of communication, Part {I}.}{A
  mathematical theory of communication, part {I}.}\BBCQ{}
\newblock
\APACjournalVolNumPages{Bell Systems Technical Journal}{27}{}{379--423}.
\PrintBackRefs{\CurrentBib}

\bibitem[\protect\citeauthoryear{%
Sokunbi%
\ \protect\BOthers{.}}{%
Sokunbi%
\ \protect\BOthers{.}}{%
{\protect\APACyear{2013}}%
}]{%
sokunbi2013}%
\APACinsertmetastar{%
sokunbi2013}%
Sokunbi, M\BPBI O.%
, Fung, W.%
, Sawlani, V.%
, Choppin, S.%
, Linden, D\BPBI E.%
\BCBL{}\ \BBA{} Thome, J.%
%
\unskip\
\newblock
\APACrefYearMonthDay{2013}{}{}.
\newblock
\BBOQ{}\APACrefatitle{Resting state {fMRI} entropy probes complexity of brain
  activity in adults with {ADHD}}{Resting state {fMRI} entropy probes
  complexity of brain activity in adults with {ADHD}}.\BBCQ{}
\newblock
\APACjournalVolNumPages{Psychiatry Research: Neuroimaging}{214}{3}{341--348}.
\PrintBackRefs{\CurrentBib}

\bibitem[\protect\citeauthoryear{%
Soler-Toscano%
, Zenil%
, Delahaye%
\BCBL{}\ \BBA{} Gauvrit%
}{%
Soler-Toscano%
\ \protect\BOthers{.}}{%
{\protect\APACyear{2013}}%
}]{%
soler2013}%
\APACinsertmetastar{%
soler2013}%
Soler-Toscano, F.%
, Zenil, H.%
, Delahaye, J\BHBI P.%
\BCBL{}\ \BBA{} Gauvrit, N.%
%
\unskip\
\newblock
\APACrefYearMonthDay{2013}{}{}.
\newblock
\BBOQ{}\APACrefatitle{Correspondence and Independence of Numerical Evaluations
  of Algorithmic Information Measures}{Correspondence and independence of
  numerical evaluations of algorithmic information measures}.\BBCQ{}
\newblock
\APACjournalVolNumPages{Computability}{2}{2}{125-140}.
\PrintBackRefs{\CurrentBib}

\bibitem[\protect\citeauthoryear{%
Soler-Toscano%
, Zenil%
, Delahaye%
\BCBL{}\ \BBA{} Gauvrit%
}{%
Soler-Toscano%
\ \protect\BOthers{.}}{%
{\protect\APACyear{2014}}%
}]{%
soler2012}%
\APACinsertmetastar{%
soler2012}%
Soler-Toscano, F.%
, Zenil, H.%
, Delahaye, J\BHBI P.%
\BCBL{}\ \BBA{} Gauvrit, N.%
%
\unskip\
\newblock
\APACrefYearMonthDay{2014}{}{}.
\newblock
\BBOQ{}\APACrefatitle{Calculating {K}olmogorov Complexity from the Output
  Frequency Distributions of Small Turing Machines}{Calculating {K}olmogorov
  complexity from the output frequency distributions of small turing
  machines}.\BBCQ{}
\newblock
\APACjournalVolNumPages{PLOS One}{9}{5}{e96223}.
\PrintBackRefs{\CurrentBib}

\bibitem[\protect\citeauthoryear{%
Solomonoff%
}{%
Solomonoff%
}{%
{\protect\APACyear{1964}}%
{\protect\APACexlab{{\protect\BCnt{1}}}}}]{%
solomonoff1964a}%
\APACinsertmetastar{%
solomonoff1964a}%
Solomonoff, R\BPBI J.%
%
\unskip\
\newblock
\APACrefYearMonthDay{1964{\protect\BCnt{1}}}{}{}.
\newblock
\BBOQ{}\APACrefatitle{A formal theory of inductive inference. {P}art {I}}{A
  formal theory of inductive inference. {P}art {I}}.\BBCQ{}
\newblock
\APACjournalVolNumPages{Information and Control}{7}{1}{1--22}.
\PrintBackRefs{\CurrentBib}

\bibitem[\protect\citeauthoryear{%
Solomonoff%
}{%
Solomonoff%
}{%
{\protect\APACyear{1964}}%
{\protect\APACexlab{{\protect\BCnt{2}}}}}]{%
solomonoff1964b}%
\APACinsertmetastar{%
solomonoff1964b}%
Solomonoff, R\BPBI J.%
%
\unskip\
\newblock
\APACrefYearMonthDay{1964{\protect\BCnt{2}}}{}{}.
\newblock
\BBOQ{}\APACrefatitle{A formal theory of inductive inference. {P}art {II}}{A
  formal theory of inductive inference. {P}art {II}}.\BBCQ{}
\newblock
\APACjournalVolNumPages{Information and Control}{7}{2}{224--254}.
\PrintBackRefs{\CurrentBib}

\bibitem[\protect\citeauthoryear{%
Takahashi%
}{%
Takahashi%
}{%
{\protect\APACyear{2013}}%
}]{%
takahashi2012}%
\APACinsertmetastar{%
takahashi2012}%
Takahashi, T.%
%
\unskip\
\newblock
\APACrefYearMonthDay{2013}{}{}.
\newblock
\BBOQ{}\APACrefatitle{Complexity of spontaneous brain activity in mental
  disorders}{Complexity of spontaneous brain activity in mental
  disorders}.\BBCQ{}
\newblock
\APACjournalVolNumPages{Progress in Neuro-Psychopharmacology and Biological
  Psychiatry}{45}{}{258--266}.
\PrintBackRefs{\CurrentBib}

\bibitem[\protect\citeauthoryear{%
Taufemback%
, Giglio%
\BCBL{}\ \BBA{} Da~Silva%
}{%
Taufemback%
\ \protect\BOthers{.}}{%
{\protect\APACyear{2011}}%
}]{%
taufemback2011}%
\APACinsertmetastar{%
taufemback2011}%
Taufemback, C.%
, Giglio, R.%
\BCBL{}\ \BBA{} Da~Silva, S.%
%
\unskip\
\newblock
\APACrefYearMonthDay{2011}{}{}.
\newblock
\BBOQ{}\APACrefatitle{Algorithmic complexity theory detects decreases in the
  relative efficiency of stock markets in the aftermath of the 2008 financial
  crisis}{Algorithmic complexity theory detects decreases in the relative
  efficiency of stock markets in the aftermath of the 2008 financial
  crisis}.\BBCQ{}
\newblock
\APACjournalVolNumPages{Economics Bulletin}{31}{2}{1631--1647}.
\PrintBackRefs{\CurrentBib}

\bibitem[\protect\citeauthoryear{%
Towse%
}{%
Towse%
}{%
{\protect\APACyear{1998}}%
}]{%
towse1998}%
\APACinsertmetastar{%
towse1998}%
Towse, J\BPBI N.%
%
\unskip\
\newblock
\APACrefYearMonthDay{1998}{}{}.
\newblock
\BBOQ{}\APACrefatitle{Analyzing human random generation behavior: A review of
  methods used and a computer program for describing performance}{Analyzing
  human random generation behavior: A review of methods used and a computer
  program for describing performance}.\BBCQ{}
\newblock
\APACjournalVolNumPages{Behavior Research Methods}{30}{4}{583-591}.
\PrintBackRefs{\CurrentBib}

\bibitem[\protect\citeauthoryear{%
Towse%
\ \BBA{} Cheshire%
}{%
Towse%
\ \BBA{} Cheshire%
}{%
{\protect\APACyear{2007}}%
}]{%
towse07}%
\APACinsertmetastar{%
towse07}%
Towse, J\BPBI N.%
\BCBT{}\ \BBA{} Cheshire, A.%
%
\unskip\
\newblock
\APACrefYearMonthDay{2007}{}{}.
\newblock
\BBOQ{}\APACrefatitle{Random number generation and working memory}{Random
  number generation and working memory}.\BBCQ{}
\newblock
\APACjournalVolNumPages{European Journal of Cognitive
  Psychology}{19}{3}{374-394}.
\PrintBackRefs{\CurrentBib}

\bibitem[\protect\citeauthoryear{%
Tversky%
\ \BBA{} Kahneman%
}{%
Tversky%
\ \BBA{} Kahneman%
}{%
{\protect\APACyear{1974}}%
}]{%
tversky1974judgment}%
\APACinsertmetastar{%
tversky1974judgment}%
Tversky, A.%
\BCBT{}\ \BBA{} Kahneman, D.%
%
\unskip\
\newblock
\APACrefYearMonthDay{1974}{}{}.
\newblock
\BBOQ{}\APACrefatitle{Judgment under uncertainty: Heuristics and
  biases}{Judgment under uncertainty: Heuristics and biases}.\BBCQ{}
\newblock
\APACjournalVolNumPages{Science}{185}{4157}{1124-1131}.
\PrintBackRefs{\CurrentBib}

\bibitem[\protect\citeauthoryear{%
Wagenaar%
}{%
Wagenaar%
}{%
{\protect\APACyear{1970}}%
}]{%
wagenaar1970}%
\APACinsertmetastar{%
wagenaar1970}%
Wagenaar, W\BPBI A.%
%
\unskip\
\newblock
\APACrefYearMonthDay{1970}{}{}.
\newblock
\BBOQ{}\APACrefatitle{Subjective randomness and the capacity to generate
  information}{Subjective randomness and the capacity to generate
  information}.\BBCQ{}
\newblock
\APACjournalVolNumPages{Acta Psychologica}{33}{}{233-242}.
\PrintBackRefs{\CurrentBib}

\bibitem[\protect\citeauthoryear{%
Wallace%
\ \BBA{} Dowe%
}{%
Wallace%
\ \BBA{} Dowe%
}{%
{\protect\APACyear{1999}}%
}]{%
wallace1999minimum}%
\APACinsertmetastar{%
wallace1999minimum}%
Wallace, C\BPBI S.%
\BCBT{}\ \BBA{} Dowe, D\BPBI L.%
%
\unskip\
\newblock
\APACrefYearMonthDay{1999}{}{}.
\newblock
\BBOQ{}\APACrefatitle{Minimum message length and {K}olmogorov
  complexity}{Minimum message length and {K}olmogorov complexity}.\BBCQ{}
\newblock
\APACjournalVolNumPages{The Computer Journal}{42}{4}{270--283}.
\PrintBackRefs{\CurrentBib}

\bibitem[\protect\citeauthoryear{%
Watanabe%
\ \protect\BOthers{.}}{%
Watanabe%
\ \protect\BOthers{.}}{%
{\protect\APACyear{2003}}%
}]{%
watanabe2003}%
\APACinsertmetastar{%
watanabe2003}%
Watanabe, T.%
, Cellucci, C.%
, Kohegyi, E.%
, Bashore, T.%
, Josiassen, R.%
, Greenbaun, N.%
\BCBL{}\ \BOthersPeriod{.}%
\unskip\
\newblock
\APACrefYearMonthDay{2003}{}{}.
\newblock
\BBOQ{}\APACrefatitle{The algorithmic complexity of multichannel {EEGs} is
  sensitive to changes in behavior}{The algorithmic complexity of multichannel
  {EEGs} is sensitive to changes in behavior}.\BBCQ{}
\newblock
\APACjournalVolNumPages{Psychophysiology}{40}{1}{77--97}.
\PrintBackRefs{\CurrentBib}

\bibitem[\protect\citeauthoryear{%
Wiegersma%
}{%
Wiegersma%
}{%
{\protect\APACyear{1984}}%
}]{%
wiegersma1984}%
\APACinsertmetastar{%
wiegersma1984}%
Wiegersma, S.%
%
\unskip\
\newblock
\APACrefYearMonthDay{1984}{}{}.
\newblock
\BBOQ{}\APACrefatitle{High-speed sequantial vocal response
  production}{High-speed sequantial vocal response production}.\BBCQ{}
\newblock
\APACjournalVolNumPages{Perceptual and Motor Skills}{59}{}{43-50}.
\PrintBackRefs{\CurrentBib}

\bibitem[\protect\citeauthoryear{%
Wilder%
, Feldman%
\BCBL{}\ \BBA{} Singh%
}{%
Wilder%
\ \protect\BOthers{.}}{%
{\protect\APACyear{2011}}%
}]{%
wilder2011}%
\APACinsertmetastar{%
wilder2011}%
Wilder, J.%
, Feldman, J.%
\BCBL{}\ \BBA{} Singh, M.%
%
\unskip\
\newblock
\APACrefYearMonthDay{2011}{}{}.
\newblock
\BBOQ{}\APACrefatitle{Contour complexity and contour detectability}{Contour
  complexity and contour detectability}.\BBCQ{}
\newblock
\APACjournalVolNumPages{Journal of Vision}{11}{11}{1044}.
\PrintBackRefs{\CurrentBib}

\bibitem[\protect\citeauthoryear{%
Williams%
\ \BBA{} Griffiths%
}{%
Williams%
\ \BBA{} Griffiths%
}{%
{\protect\APACyear{2013}}%
}]{%
williams_why_2013}%
\APACinsertmetastar{%
williams_why_2013}%
Williams, J\BPBI J.%
\BCBT{}\ \BBA{} Griffiths, T\BPBI L.%
%
\unskip\
\newblock
\APACrefYearMonthDay{2013}{}{}.
\newblock
\BBOQ{}\APACrefatitle{Why are people bad at detecting randomness? A statistical
  argument}{Why are people bad at detecting randomness? a statistical
  argument}.\BBCQ{}
\newblock
\APACjournalVolNumPages{Journal of Experimental Psychology: Learning, Memory,
  and Cognition}{39}{5}{1473--1490}.
\PrintBackRefs{\CurrentBib}

\bibitem[\protect\citeauthoryear{%
Yagil%
}{%
Yagil%
}{%
{\protect\APACyear{2009}}%
}]{%
yagil2009}%
\APACinsertmetastar{%
yagil2009}%
Yagil, G.%
%
\unskip\
\newblock
\APACrefYearMonthDay{2009}{}{}.
\newblock
\BBOQ{}\APACrefatitle{The structural complexity of DNA templatesImplications
  on cellular complexity}{The structural complexity of dna
  templatesimplications on cellular complexity}.\BBCQ{}
\newblock
\APACjournalVolNumPages{Journal of Theoretical Biology}{259}{3}{621--627}.
\PrintBackRefs{\CurrentBib}

\bibitem[\protect\citeauthoryear{%
Yamada%
, Kawabe%
\BCBL{}\ \BBA{} Miyazaki%
}{%
Yamada%
\ \protect\BOthers{.}}{%
{\protect\APACyear{2013}}%
}]{%
yamada_pattern_2013}%
\APACinsertmetastar{%
yamada_pattern_2013}%
Yamada, Y.%
, Kawabe, T.%
\BCBL{}\ \BBA{} Miyazaki, M.%
%
\unskip\
\newblock
\APACrefYearMonthDay{2013}{}{}.
\newblock
\BBOQ{}\APACrefatitle{Pattern randomness aftereffect}{Pattern randomness
  aftereffect}.\BBCQ{}
\newblock
\APACjournalVolNumPages{Scientific Reports}{3}{}{}.
\PrintBackRefs{\CurrentBib}

\bibitem[\protect\citeauthoryear{%
Yang%
\ \BBA{} Tsai%
}{%
Yang%
\ \BBA{} Tsai%
}{%
{\protect\APACyear{2012}}%
}]{%
yang2012}%
\APACinsertmetastar{%
yang2012}%
Yang, A\BPBI C.%
\BCBT{}\ \BBA{} Tsai, S\BHBI J.%
%
\unskip\
\newblock
\APACrefYearMonthDay{2012}{}{}.
\newblock
\BBOQ{}\APACrefatitle{Is mental illness complex? {F}rom behavior to brain}{Is
  mental illness complex? {F}rom behavior to brain}.\BBCQ{}
\newblock
\APACjournalVolNumPages{Progress in Neuro-Psychopharmacology and Biological
  Psychiatry}{45}{}{253--257}.
\PrintBackRefs{\CurrentBib}

\bibitem[\protect\citeauthoryear{%
Zabelina%
, Robinson%
, Council%
\BCBL{}\ \BBA{} Bresin%
}{%
Zabelina%
\ \protect\BOthers{.}}{%
{\protect\APACyear{2012}}%
}]{%
zabelina2012}%
\APACinsertmetastar{%
zabelina2012}%
Zabelina, D\BPBI L.%
, Robinson, M\BPBI D.%
, Council, J\BPBI R.%
\BCBL{}\ \BBA{} Bresin, K.%
%
\unskip\
\newblock
\APACrefYearMonthDay{2012}{}{}.
\newblock
\BBOQ{}\APACrefatitle{Patterning and nonpatterning in creative cognition:
  Insights from performance in a random number generation task.}{Patterning and
  nonpatterning in creative cognition: Insights from performance in a random
  number generation task.}\BBCQ{}
\newblock
\APACjournalVolNumPages{Psychology of Aesthetics, Creativity, and the
  Arts}{6}{2}{137--145}.
\PrintBackRefs{\CurrentBib}

\bibitem[\protect\citeauthoryear{%
Zenil%
}{%
Zenil%
}{%
{\protect\APACyear{2011}}%
{\protect\APACexlab{{\protect\BCnt{1}}}}}]{%
zenil2011book}%
\APACinsertmetastar{%
zenil2011book}%
Zenil, H.%
%
\unskip\
\newblock
\APACrefYear{2011{\protect\BCnt{1}}}.
\newblock
\APACrefbtitle{Randomness Through Computation: Some Answers, More
  Questions}{Randomness through computation: Some answers, more questions}.
\newblock
\APACaddressPublisher{}{World Scientific}.
\PrintBackRefs{\CurrentBib}

\bibitem[\protect\citeauthoryear{%
Zenil%
}{%
Zenil%
}{%
{\protect\APACyear{2011}}%
{\protect\APACexlab{{\protect\BCnt{2}}}}}]{%
zenil2011thesis}%
\APACinsertmetastar{%
zenil2011thesis}%
Zenil, H.%
%
\unskip\
\newblock
\APACrefYear{2011{\protect\BCnt{2}}}.
\newblock
\APACrefbtitle{Une approche exp{\'e}rimentale de la th{\'e}orie algorithmique
  de la complexit{\'e}}{Une approche exp{\'e}rimentale de la th{\'e}orie
  algorithmique de la complexit{\'e}}.
\newblock
\BUPhD, Universidad de Buenos Aires.
\PrintBackRefs{\CurrentBib}

\bibitem[\protect\citeauthoryear{%
Zenil%
\ \BBA{} Delahaye%
}{%
Zenil%
\ \BBA{} Delahaye%
}{%
{\protect\APACyear{2010}}%
}]{%
AlgNatZenil}%
\APACinsertmetastar{%
AlgNatZenil}%
Zenil, H.%
\BCBT{}\ \BBA{} Delahaye, J\BHBI P.%
%
\unskip\
\newblock
\APACrefYearMonthDay{2010}{}{}.
\newblock
\BBOQ{}\APACrefatitle{ON THE ALGORITHMIC NATURE OF THE WORLD}{On the
  algorithmic nature of the world}.\BBCQ{}
\newblock
\BIn{} G.~Dodig-Crnkovic\ \BBA{} M.~Burgin\ (\BEDS), \APACrefbtitle{Information
  and Computation}{Information and computation}\ (\BPG~477-496).
\newblock
\APACaddressPublisher{}{World Scientific}.
\PrintBackRefs{\CurrentBib}

\bibitem[\protect\citeauthoryear{%
Zenil%
\ \BBA{} Delahaye%
}{%
Zenil%
\ \BBA{} Delahaye%
}{%
{\protect\APACyear{2011}}%
}]{%
zenil2011}%
\APACinsertmetastar{%
zenil2011}%
Zenil, H.%
\BCBT{}\ \BBA{} Delahaye, J\BHBI P.%
%
\unskip\
\newblock
\APACrefYearMonthDay{2011}{}{}.
\newblock
\BBOQ{}\APACrefatitle{An algorithmic information theoretic approach to the
  behaviour of financial markets}{An algorithmic information theoretic approach
  to the behaviour of financial markets}.\BBCQ{}
\newblock
\APACjournalVolNumPages{Journal of Economic Surveys}{25}{3}{431--463}.
\PrintBackRefs{\CurrentBib}

\bibitem[\protect\citeauthoryear{%
Zenil%
, Soler{-}Toscano%
, Delahaye%
\BCBL{}\ \BBA{} Gauvrit%
}{%
Zenil%
\ \protect\BOthers{.}}{%
{\protect\APACyear{2012}}%
}]{%
Kolmo2D}%
\APACinsertmetastar{%
Kolmo2D}%
Zenil, H.%
, Soler{-}Toscano, F.%
, Delahaye, J.%
\BCBL{}\ \BBA{} Gauvrit, N.%
%
\unskip\
\newblock
\APACrefYearMonthDay{2012}{}{}.
\newblock
\BBOQ{}\APACrefatitle{Two-Dimensional {K}olmogorov Complexity and Validation of
  the Coding Theorem Method by Compressibility}{Two-dimensional {K}olmogorov
  complexity and validation of the coding theorem method by
  compressibility}.\BBCQ{}
\newblock
\APACjournalVolNumPages{CoRR}{abs/1212.6745}{}{}.
\newblock
 \begin{APACrefURL} \url{http://arxiv.org/abs/1212.6745} \end{APACrefURL}
\PrintBackRefs{\CurrentBib}

\end{thebibliography}
\end{document}